\documentclass[12pt,preprint]{aastex}
\def\lax {\ifmmode{_<\atop^{\sim}}\else{${_<\atop^{\sim}}$}\fi}  
\def\gax {\ifmmode{_>\atop^{\sim}}\else{${_>\atop^{\sim}}$}\fi}  
\def\gtorder{\mathrel{\raise.3ex\hbox{$>$}\mkern-14mu
             \lower0.6ex\hbox{$\sim$}}}

\def\cm2{cm$^{-2}$}
\def\s1{s$^{-1}$}

\begin{document}

\title{How to distinguish white dwarf and neutron star X-ray binaries during their X-ray outbursts?}

\author{Lev Titarchuk\altaffilmark{1} and Elena Seifina\altaffilmark{2}}

\altaffiltext{1}{Dipartimento di Fisica, Universit\`a di Ferrara, Via Saragat 1, I-44100 Ferrara, Italy, email:titarchuk@fe.infn.it; George Mason University Fairfax, VA 22030;   
Goddard Space Flight Center, NASA,  code 663, Greenbelt  
MD 20770, USA; email:lev@milkyway.gsfc.nasa.gov, USA}
\altaffiltext{2}{Moscow M.V.~Lomonosov State University/Sternberg Astronomical Institute, Universitetsky 
Prospect 13, Moscow, 119992, Russia; seif@sai.msu.ru}

\begin{abstract}
We present spectral 
signatures of neutron stars (NSs) and white dwarfs (WDs) hosted in accreting X-ray binaries that can be easily identified in X-ray observations. We perform spectral and timing analysis of 4U~1636--53 and SS~Cygni, as typical representatives of such NS and WD binaries, based on their X-ray observations by {\it RXTE}, ASCA, {\it Suzaku} and {\it Beppo}SAX uising {\it Comptonization} spectral model. As a result, we formulate a criterion that makes it easy to distinguish NS from WD in such binaries: NS X-rays exhibits quasi-stable behavior with the index $\Gamma\to2$ and is characterized by quasi periodic oscillations (QPOs) at $\nu_{QPO} >0.5$~Hz, although WD X-rays is stable with $\Gamma \to1.85$ and is accompanied by QPOs at $\nu_{QPO}<0.05$~Hz during source outbursts. In addition,  we revealed that in 4U~1636--53 the mHz QPOs anti-correlate with the plasma temperature, $T_e$ of Compton cloud (or the corona around a NS. This allowed us to associate mHz-QPOs  with the corona dynamics during outburst cycle. The above index effect, now well established for 4U~1636--53 and SS~Cygni  using extensive observations, has previously been found in other low-mass X-ray NS and WD binaries and agrees well with the criterion for distinguishing NSs and WDs presented here. 
\end{abstract}


\section{Introduction}

Compact objects – black holes, neutron stars (NSs) and white dwarfs (WDs) represent a unique opportunity to study stars at its late evolution stages and, thus, to study the properties of matter in superdense states that are not found in terrestrial conditions. These objects have some differences, both observational and theoretical. For example, NSs and WDs, in contrast to black holes, have a hard surface. The NS and WD themselves differ in size and physical parameters. NSs and WDs, which are hosted in low-mass X-ray binaries (LMXBs), sometimes accreting matter from a companion star, causing X-ray outbursts. These episodes give their clear observational signatures of NSs and WDs and thus help us to formulate similarities and differences in their observational manifestations. In X-ray outbursts, they follow a similar evolution, demonstrating different spectral states, probably associated with different accretion regimes. Of these spectral states, two of the most important can be distinguished: the low-hard states (LHS) and the high-soft states (HSS). It is believed that Compton scattering by hot electrons predominates in the hard state; and the soft state is associated with thermal radiation, for an example from the accretion disk and a NS surface in the NS case. An evolution of such sources during X-ray outbursts follows a {\it q}-shaped counterclockwise motion trajectory in terms of a hardness-intensity diagram (HID). Outbursts can last from a few weeks to many months before eventually returning to a low-luminosity dormant state. In addition to HID, low-frequency quasi peridic oscillations (LF-QPOs) have proven to be a useful tool for tracking spectral states and transitions between them. Recently, \cite{Sonbas22}  introduced a ``Minimum Time Scale" (MTS),  a timing property, to track spectral transitions with an intensity variability diagram that plots the source count rate as a function of MTS as the source undergoes spectral changes during an outburst. The MTS represents the time scale associated with the shortest time response in the light curve, or equivalently, the highest frequency component of the signal above the Poisson noise in the power spectra \cite{Mohamed21}. 

%
%
\begin{table}
\centering
\caption{Basic parameters of 4U~1636--53 and SS~Cygni binaries.}\label{tab:parameters_binaries} 
 \begin{tabular}{lcccl}
 \hline\hline                        
  Parameter               &     4U~1636--53     &   SS~Cygni      \\
      \hline
Class of primary         & NS, $atoll$            & WD \\
Mass of an X-ray star, M$_{\odot}$ & $\sim$1.4~\cite{Giles02}              &1.19 $\pm$ 0.02~\cite{Friend90},   0.81$ \pm$ 0.19~\cite{Bitner07}\\
Spectral type  of secondary    & ...                         & K5/5 \\
Mass of an optical star, M$_{\odot}$ & $\sim$0.4              &0.70 $\pm$ 0.02~\cite{Friend90} \\
Orbital inclination,  $i$,  deg                   & 70~\cite{Sanna13,Casares06,Frank87}                          &37 $\pm$ 5~, 45--56~\cite{Bitner07}  \\
Distance, 
pc       & (6.0$\pm$0.5)$\times 10^{3}$~\cite{Ritter_Kolb03}               & 166 $\pm$ 12~\cite{Galloway20} \\
Orbital period, 
hr & 3.8~\cite{smale_mukai88,van_Paradijs90}                          &6.603~\cite{Harrison04} \\
 \hline                                             
\end{tabular}
\end{table}

%
%
\begin{figure}
\centering
\includegraphics[width=0.8\textwidth]{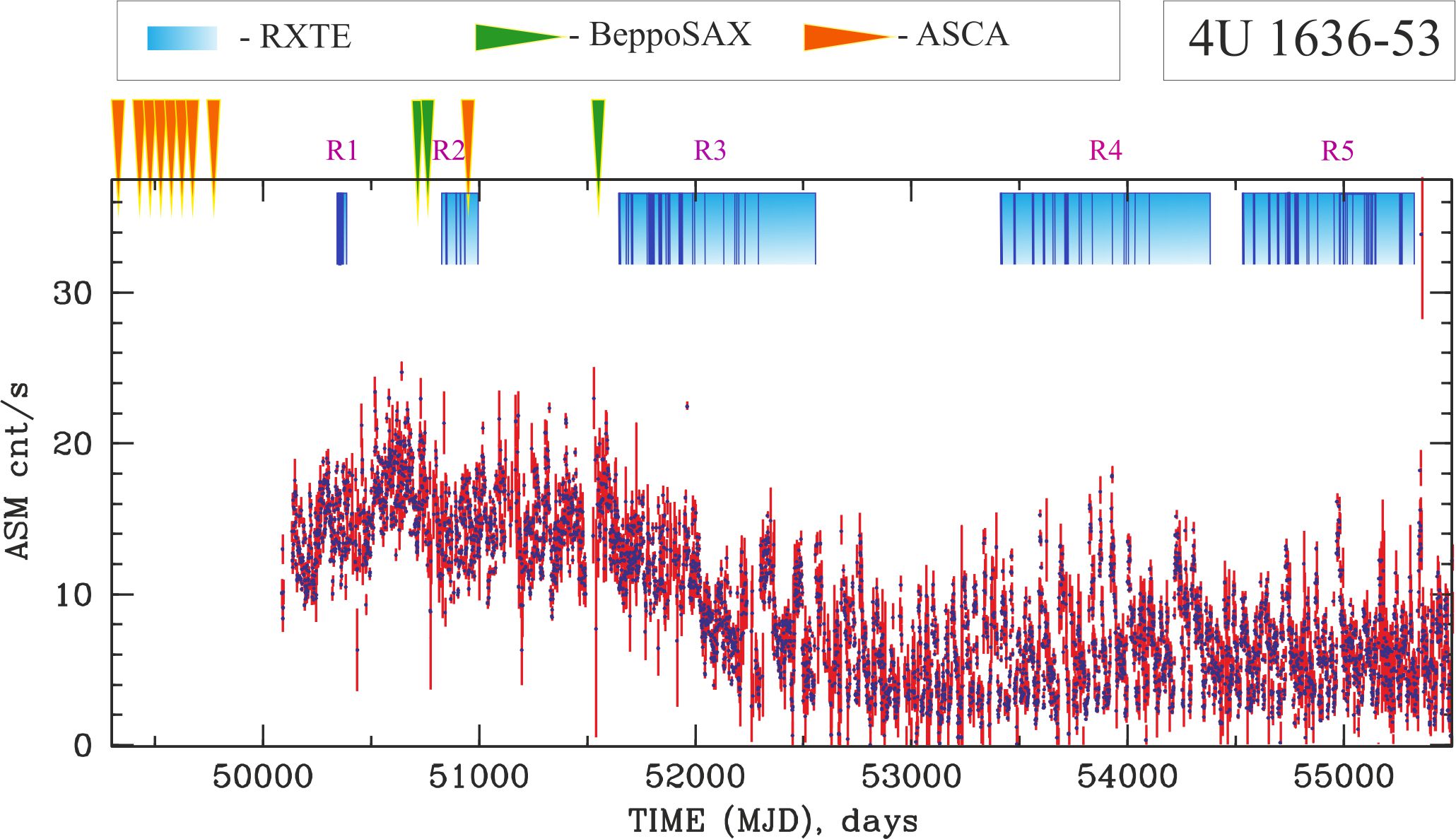}
\caption{ Evolution of ASM/{\it RXTE} count rate during  1996 -- 2010 observations of 4U~1636--53.
{\it Green} and {\it orange} triangles show {\it Beppo}SAX and ASCA data, respectively, listed in Table~\ref{tab:table_SAX+ASCA_1636}.  
{\it Bright blue} boxes 
are related to  the {\it RXTE} data sets listed in Table~\ref{tab:par_RXTE_1636}. 
}
\label{variability_96-10}
\end{figure}

A task of searching/formulating distinctive features of NSs and WDs that are easily recognizable in observations has become urgent. 
In this paper, we study observational differences in terms of X-ray spectra using  samples of two bright representatives of these classes, 4U~1636--53 and SS~Cygni, during their X-ray outbursts. 
The low mass X-ray binary (LMXB) 4U~1636--53  is a member of the $atoll$ class~\cite{hasinger89}. It contains a 
NS accreting matter from a companion star of mass $\sim$0.4~M$_{\odot}$ (see Table~\ref{tab:parameters_binaries}). This source has been extensively observed over the past five decades and is considered as a persistent X-ray source~\cite{liu07}. Its optical counterpart (V801 Ara) revealed its 3.8~hr orbital period~(see, e.g.  \cite{smale_mukai88} and \cite{van_Paradijs90}) as photometric variations, but the period has also been detected spectroscopically~\cite{Casares06}.

4U~1636--53 shows bursts of thermonuclear combustion on the surface of a star, which immediately classifies it as a reliable NS source~\cite{stroh98}. The distance to 4U~1636--536 was determined to be 6.0$\pm$0.5 kpc, assuming that its brightest outbursts with an expansion of the photospheric radius  caused by pure helium \cite{Galloway06,Galloway18}, which is consistent with the measured value of 4.42$^{+3,08}_{-1.63}$ kpc by {\it Gaia} \cite{Galloway20}.  Previous studies suggest this source has a high inclination, however a lack of dips in the X-ray light curve limits the inclination $\sim 70^{\circ}$  \cite{Sanna13,Casares06,Frank87}. X-ray observations with the  All Sky Monitor (ASM) onboard {\it RXTE} 
shown that light curve of 4U~1636--53  demonstrates two kinds of long-term segments with different mean flux levels. Up to 2000, the ASM mean count-rate was significantly higher than that after 2001  (see Fig.~\ref{variability_96-10}). During 2001, the ASM recorded an obvious, gradual decline of the X-ray flux.
Within a year, it began to exhibit a substantial quasi-periodic, long-term variability of 30--40 days~\cite{Shih05}.
 This variability   is consistent with  temporal and energy changes that occur between two states of {\it atoll} sources: the {\it island} (hard) and {\it banana} (soft) states~\cite{hasinger89}. Furthermore, \cite{Shih05} also reported that this variation was present in both X-ray soft (2--12 keV, {\it RXTE}/ASM) and hard (20--100 keV, INTEGRAL/IBIS) energy bands, but that these ones were anti-correlated.  Such a clear energy-dependent pattern is explained by   an accretion disc instability.

%
%
\begin{table*}
 \caption{The list of {\it Beppo}SAX and ASCA observations of 4U~1636--53  used in our analysis.}
   \label{tab:table_SAX+ASCA_1636}
 \begin{tabular}{llcccc}
 \hline\hline  
Set & Satellite & Obs. ID& Start time (UT)  & End time (UT) &MJD interval \\
      \hline
B1 & BS & 20168003& 1998 Feb 28 22:40:24 & 1998 Mar 1 12:50:17 &50873.0--50873.5~\cite{fiocchi06} \\
B2 & BS & 20537002& 1998 Feb 24 19:20:13 & 1998 Feb 24 15:07:18 &50868.8--50869.6~\cite{fiocchi06} \\
B3 & BS & 20836001& 2000 Feb 15 23:52:07 & 2000 Feb 17 01:11:09 &51589.9--51591.1~\cite{fiocchi06} \\
      \hline
A1 & ASCA           & 40026000& 1993 Aug 9 10:31:22 & 1993 Aug 10 15:53:50 &49208.4--49209.7\\
A2 & ASCA           & 42004000& 1994 Aug 31 06:43:44 & 1994 Sep 1 10:40:48 &49595.2--49596.4\\
A3 & ASCA           & 42004010& 1994 Sep 9 04:34:17 & 1994 Sep 10 06:10:35 &49604.2--49605.3\\
A4 & ASCA           & 42004020& 1994 Sep 5 12:44:18 & 1994 Sep 5 17:11:54 &49600.5--49605.7\\
A5 & ASCA           & 42004030& 1994 Sep 13 01:05:00 & 1994 Sep 13 21:23:40 &49608.0--49608.9\\
A6 & ASCA           & 42004040& 1994 Sep 7 17:32:44 & 1994 Sep 8 18:46:54 &49602.7--49603.8\\
A7 & ASCA           & 42004050& 1994 Sep 6 11:41:54 & 1994 Sep 6 17:09:18 &49601.5--49601.7\\
A8 & ASCA           & 42004060& 1995 Mar 20 06:32:58 & 1995 Mar 21 03:24:10 &49796.3--49797.1\\
A9 & ASCA           & 46011000& 1998 Aug 19 09:23:04 & 1998 Aug 20 06:52:08 &51044.3--51045.3~\cite{CK00} \\
      \hline
      \end{tabular}
\end{table*}

Belloni et al. (2007) revealed the dependence of timing pattern on the mean X-ray luminosity, sampling effect and spectral hardness of 4U~1636--35 as well as provided further evidence regarding the nature of the long-term variability of this source~\cite{belloni07}. Twenty years after the discovery of kHz  quasi-periodic oscillations (QPOs) in NS low-mass X-ray binaries  \cite{Strohmayer96,van_der_Klis96},  a   pretty convincing explanation of the origin of this QPO phenomenon remains a challenge  \cite{Mendez06}.  
%
Millihertz quasi-periodic oscillations (mHz QPOs)~\cite{Revnivtsev01,Altamirano08,Lyu15}
and their harmonics~\cite{Fei21} 
were found in 4U~1636--53, which were immediately attributed to nuclear combustion processes on the surface of a NS. However, Heger et al. (2007) showed that stable nuclear burning of helium on the surface of a neutron star can lead to mHz-QPO only in the case of the Eddington accretion regime~\cite{Heger07}. Such a regime corresponds to an accretion rate an order of magnitude higher than the average accretion rate determined from observations. To address this inconsistency, Heger et al. (2007) suggested that the local accretion rate in the burning layer, where mHz QPOs originate, may be higher than the global rate. Regarding variation in chemical composition as a reason for the formation of mHz-QPOs, Keek et al. (2014) found that mHz-QPOs cannot be initiated at observed accretion rates by changing only the chemical composition and nuclear reaction rate~\cite{Keek14}.

Altamirano et al. (2008) found that the frequency of mHz-QPOs in 4U~1636--53 systematically decreased before the type I X-ray burst, and then the QPOs disappeared when the burst occurred~\cite{Altamirano08}. Recently, the same behavior was found in another mHz QPO source, LMXB EXO~0748--676~\cite{Mancuso19}. The frequency drift behavior of QPOs in mHz indicates that QPOs in mHz are closely related to nuclear burning on the surface of a neutron star. In addition, Linares et al. (2012) reported a smooth evolution between their Hz QPOs and X-ray bursts in IGR~J17480--2446: as the accretion rate increased, the bursts gradually turned into mHz-QPOs, and vice versa~\cite{Linares12}. Recently, other properties of the mHz-QPO have been investigated. Lyu et al. (2015) found that the mHz-QPO frequency does not have a significant correlation with the surface temperature of a NS in 4U~1636--53, contrary to model predictions~\cite{Lyu15}.

%
%

\begin{table}
  \centering 
 \caption{The 
sets of  {\it RXTE} observation of 4U~1636--53.}
 \begin{tabular}{lllllc}
      \hline\hline
Set  & Dates, MJD & Obs. ID&  Dates UT \\
 \hline
R1  &    50446--50503 & 10088                       & Dec 29, 1996 -- Feb 24, 1997 \\
R2  &    50869--51118 & 30053~\cite{CK00}, 40028, 40031          & Feb 25, 1998 --  June 15, 1999 \\
R3  &    51852--52646 & 50030, 60032            & Nov 4, 2000 --    Jan 7, 2003    \\
R4  &    53606--54371 & 91027~\cite{Lyu14}, 91152~\cite{belloni07}, 93091  & Aug 29, 2005 --  Sep 28, 2007 \\
R5  &    54523--55336 & 93091~\cite{Lyu14}, 93082, 94310~\cite{Lyu14}, 94437          & Feb 27, 2008 --   May 20, 2010\\
      \hline
      \end{tabular}
    \label{tab:par_RXTE_1636}
\end{table}

In addition, Stiele et al. (2016) studied the phase-resolved energy spectra of the mHz QPOs in 4U 1636–53 and concluded that the QPOs were not associated with modulations of the neutron star surface temperature~\cite{Stiele16}. In contrast, Strohmayer et al. (2018) recently found that, in GS 1826–238, the mHz fluctuations are consistent with the change in the black-body temperature on the surface of the neutron star~\cite{Strohmayer18}. To date, the question of the origin of the mHz QPO in NS remains open. Therefore, in this paper, we investigated this problem within the framework of the transition layer (TL) model in order to search for other possible interpretations.

%
%

\begin{figure}
\centering
\includegraphics[scale=0.6,angle=0]{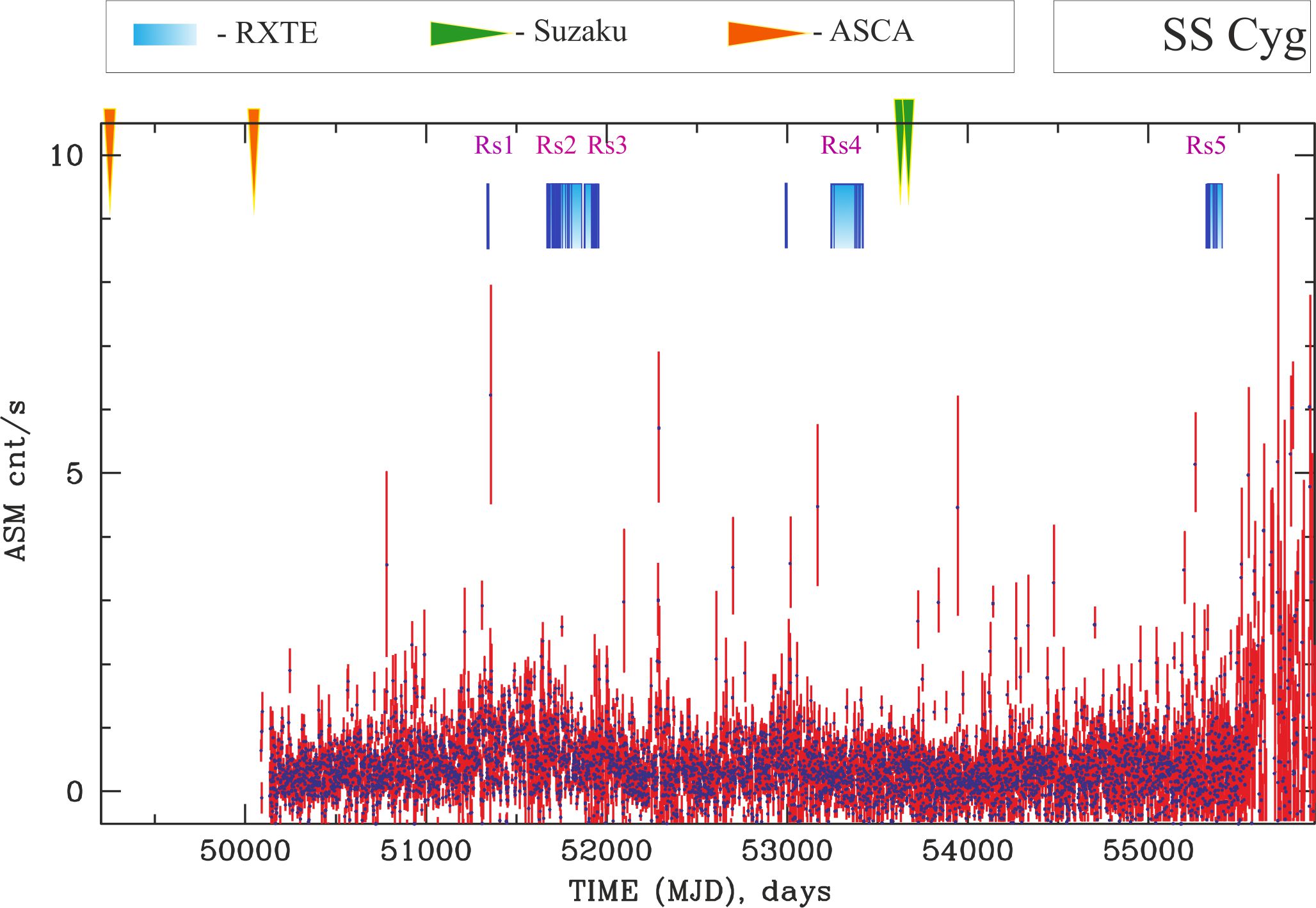}
\caption{ Evolution of ASM/{\it RXTE} count rate during  1996 -- 2010 observations of SS~Cygni. 
{\it Blue} vertical strips ({\it on top of the panel}) indicate to time  for the {\it RXTE} 
observations. 
{\it Green} triangles show {\it Suzaku} observations and {\it orange} triangles indicate {\it ASCA} observations of SS~Cygni 
(see Table~\ref{tab:table_suzaku_ss_cyg}).
}
\label{variability_96-10_ss_cyg}
\end{figure}

It is commonly supposed that the X-ray spectrum of LMXBs  is made of a soft thermal component because of  the NS/WD surface and the  accretion disk while  a hard component is formed due to the  Comptonization   of soft thermal  NS/WD and disk soft photons in a corona consisting of energetic electrons \cite{Barret01,Lin07}.  
In cataclysmic variables (CVs) with matter transfer, accretion disks (ADs) are usually too very cold  ($kT\ll 1$ keV) to emit X-rays \cite{Lewin+vanderKlis06}. The main X-ray source has been associated with the TL, the region between the spiraling AD and the surface of the slower rotating disk, where a most part of the gravitational energy is emitted. The physical characteristics of the TL and described in terms of the observed soft and hard X-ray spectral emission \cite{TSS14}.

SS~Cygni binary hosts a WD with a mass of 1.19$\pm$0.02 M$_{\odot}$ (see Table~\ref{tab:parameters_binaries}), a secondary star with a mass of 0.704$\pm$0.002 M$_{\odot}$ \cite{Friend90}, and a K5/5 spectral type \cite{Ritter_Kolb03}. This binary is distanced at 166$\pm$12~pc  \cite{Harrison04}  and is characterized by an inclination $i$ of 37$^{\circ}\pm$ 5$^{\circ}$ \cite{Ritter_Kolb03} and an orbital period oforbital rotation 6.603~hr \cite{Harrison04}. This source is one of the well studied in terms of outburst states, as it shows outburst in optical  band every $\sim$50 days \cite{Mauche+Robinson01,Lewin+vanderKlis06}. 

Lewin and van der Klis (2006)  revealed a reflection component in the  4U~1636--53 spectrum during both the quiescent and active states, with a larger contribution in the active phases  \cite{Done_Osborne97,Lewin+vanderKlis06}. Mukai et al. (2003) analyzed the $Chandra$ spectrum in quiescence in frame of MKCFLOW model and found that the plasma temperature $T_{e}$ increases to 80 keV~\cite{Mukai03}.  Okada et al (2008) also analyzed the {\it Chandra} observations of SS~Cygni and argued that in quiescence, the spectra are dominated by the H-like iron K$_{\alpha}$-lines, whereas at outburst the He-like iron lines are as intense as the H-like irom lines~\cite{Okada08}.  As a result, they 
formulated a scenario for the formation of radiation in SS~Cygni: the observed K$_{\alpha}$-emission lines from O to Si originate in the transition region between the optically thick AD and the optically thin TL region, where the Keplerian motion gradually transforms into thermal motion due to hard friction. They constrained the temperature of the TL from 1.6 keV (outburst) to 4.1 keV (quiescence) using the {\it bremss} model. 

A maximum temperature of the plasma $T_{e}\sim$ 20.4 keV in quiescence and 6 keV in outburst revealed by \cite{Ishida09} during a Suzaku 
observation of this source on the 2005 November in the quiescent and outburst states using the  CVMEKAL  model. \cite{Byckling10} also explored these data but obtained different temperatures (10--42 keV)  using the MEKAL and KCFLOW models in the quiescence. They also found iron line with energy of 6.4 keV in  SS~Cygni with the equivalent width $EW_{6.4}$ of 75$\pm$4 eV. Xu et al. (2016)
analyzed {\it Suzaku} data of SS~Cygni in 2005 using the CEVMKL model and found a maximum plasma temperature $T_e\sim$ 20.8 keV (in the quiescence state) and minimal plasma temperature of 5.8$\pm$0.3 keV (in the outburst state)~\cite{Xu16}.  Finally, Maiolino et al. (2020) demonstrated a presence of strong emission lines at energies of 1--1.5 keV and 6--7 keV against the Comptonization continuum and a plasma temperature variation from $T_e\sim$ 5 keV to 25 keV in the outburst and the quiescence, respectively using the Chandra HETG/ACIS observation of SS~Cygni using the thermal{\tt COMPTT} and {\tt Gaussian} models~\cite{Maiolino20}.

\begin{table*}
  \centering 
 \caption{The list of {\it Suzaku} and ASCA observations of SS~Cyg  used in our analysis.}
   \label{tab:table_suzaku_ss_cyg}
 \begin{tabular}{lcccccc}
 \hline\hline  
Set &Satellite &Obs. ID& Start time (UT)  & End time (UT) &MJD interval \\
      \hline
S1 & Suzaku & 400006010& 2005 Nov 2 01:12:32 & 2005 Nov 2 23:39:24 &53676.0--53676.9~\cite{Ishida09}   \\
S2 & Suzaku &400007010& 2005 Nov 18 14:33:32 & 2005 Nov 19 20:45:08 &53692.6--53692.8~\cite{Ishida09} \\
S3 & Suzaku &109015010& 2014 Dec 19 06:23:57 & 2014 Dec 19 23:56:18 &57010.2--57010.8        \\
      \hline
As1 & ASCA &30001000& 1993 May 26 21:06:16 & 1993 May 27 18:22:46 &49133.8--49134.6~\cite{Done_Osborne97}  \\
As2 & ASCA &30004000& 1995 Nov 27 04:01:52 & 1995 Nov 27 17:50:21 &50048.1--50048.5~\cite{Done_Osborne97} \\
      \hline
      \end{tabular}
\end{table*}

\begin{table}
  \centering 
 \caption{The 
groups of {\it RXTE} observation of SS~Cyg.}
 \begin{tabular}{lllllc}
      \hline\hline
Set  & Dates, MJD & Obs. ID&  Dates UT\\
       &                 &            &           \\
 \hline
Rs1  &    51337--51350 & 40012~\cite{balman12}   & Jun 8 -- 21, 1999 \\
Rs2  &    51609--51697 & 50011~\cite{balman12}        & March 6 -- Jun 2, 2000 \\
Rs3  &    51780--51926 & 50012~\cite{balman12,Maiolino20} & Aug 24, 2000 -- Jan 17, 2001\\
Rs4  &    53267--53395 & 90007        & Sep 19, 2004 -- Jan 25, 2005\\
Rs5  &    55312--55328 & 95421~\cite{balman12}       & Apr 26 -- May 12, 2010 \\
      \hline
      \end{tabular}
    \label{tab:rxte_ss_cyg}
\end{table}

Recently, Kimura et al. (2021)  pointed to an unusual failure of SS~Cygni outburst activity that occurred in 2021 in the form of a long quiescent state with reduced luminosity~\cite{Kimura21}. A similar phenomenon has already been observed earlier in Z~Can-type DNes. They 
analyzed the source spectra with the {\it tbabs*(reflect*cevmkl + Gaussian)} model using NICER and NuStar observation. They showed that during this standstill phase the observed gradual increase in optical and X-ray emissions is due to contributions from the inner part (TL) of the accretion disk, even in such a quiescent state.

Here we develop the comparative analysis for 4U~1636--53 and SS~Cygni using the {\it RXTE}, ASCA, {\it Beppo}SAX  
 and  $Suzaku$ observations. 
 In \S 2 we present the list of observations used in our data analysis while 
in \S 3 we provide the details of X-ray spectral analysis.  We analyze an evolution of X-ray spectral and timing  properties during the state transition in \S 4.   In \S 5 we present  a description of the spectral models used for fitting 
these data. 
In  \S  6 we  discuss  the  main results of the paper. In \S 7 we present our final conclusions.

\section{DATA SELECTION \label{data}}


{\bf BeppoSAX data}. We used {\it Beppo}SAX (BS) data of 4U~1636--53, which were obtained on February 24--28, 1998 and February 15--17, 2000. In Table~\ref{tab:table_SAX+ASCA_1636} (top part) we present the log of the BS 
observation analyzed  in this paper. We analyzed  the related BS spectral distributions obtained  from three BS 
Narrow Field Instruments (NFIs): the Low Energy Concentrator Spectrometer (LECS, \cite{parmar97}) for 0.3 -- 4 keV, the Medium Energy Concentrator Spectrometer (MECS, \cite{boel97}) for 1.8 -- 10 keV and the Phoswich Detection System (PDS, \cite{fron97}) for 15 -- 60 keV.  The SAXDAS data analysis software was employed in order to analyze this data set. 
We made this investigation where  the response matrix  is definitely found. Both LECS and MECS spectra were accumulated in circular  
regions of 8${\tt '}$ radius. We renormalized the  LECS  data  using  MECS as a base template.  We treat normalizations  as fit-free parameters  and 1   for   NFIs and MECS, correspondingly. We checked after that this fitting procedure and if normalizations   were in a standard range for each instrument. We also re-binned the data  in order to increase their  significance. For example, for the LECS data  the binning factor changes with energy  using  re-binnig template files in GRPPHA of XSPEC  (see Cookbook for the BS-NFI spectral  analysis). We  applied  a linear binning factor 2 (for which resulting bin width is 1 keV for the PDS spectra and  with systematic errors 1\%).  

{\bf ASCA data}. {\it ASCA} implemented observations of  4U~1636--53 on 1993---1995 and 1998, and also SS~Cygni on 1993 and 1995. In Tables~\ref{tab:table_SAX+ASCA_1636} and \ref{tab:table_suzaku_ss_cyg} (bottom parts)  we summarized the start time, end time, and the MJD interval  for 4U~1636--53 and SS~Cygni, respectively, indicated by {\it orange} triangles in top of Figs.~\ref{variability_96-10} and \ref{variability_96-10_ss_cyg}. 
The {\it ASCA} instrumentation and  data analysis was described  in \cite{Tanaka94}.  
The {\it ASCA} data were screened using the ftool {\tt ascascreen} and the standard screening criteria. The spectrum for both sources were extracted using spatial regions with a diameter of 
4${\tt '}$ 
centered on the nominal position of the source. The background was extracted using source-free regions.   
We implemented the data re-binning  in order to improve statistics  (> 20 counts per spectral bin) and  also  applied  the $\chi^2$-technique. The 0.6 -- 10 keV and 0.8 -- 10 keV ranges  were used to fit the solid and gas imaging spectrometer data, respectively. The background was extracted using  source-free regions.

{\bf Suzaku data}. {\it  Suzaku} observed SS~Cygni  in the quiescence (November 2, 2005 and November 27, 2014) and  at the outburst  (November 18, 2005). Table~\ref{tab:table_suzaku_ss_cyg} (top part) shows the start--end times, and the MJD interval duration,  indicated by {\it green} triangles in top of Fig.~\ref{variability_96-10_ss_cyg}. Mitsuda et al. (2007) and Koyama et al. (2007) present a description of the {\it Suzaku} instrumentation \cite{Mitsuda07,Koyama07}. We used observations of SS~Cygni, performed by a focal X-ray CCD camera of the X-ray Imaging Spectrometer [XIS, 0.3--12~keV energy range] and processed them with  the {\it Suzaku} data processing {\tt pipeline}. 
The {\tt HEASOFT software package} of version 6.25  and the {\it Suzaku} Data Reduction Guide\footnote{http://heasarc.gsfc.nasa.gov/docs/suzaku/analysis/} were applied in order to make the data reduction. The source spectra  were extracted using spatial regions within the 3.51${\tt '}$-radius circle centered on the nominal position of SS~Cygni  [$\alpha=21^{h}42^{m}42^s.80$, $\delta=+43^{\circ} 35{\tt '} 09{\tt ''}.8$, J2000.0, 
while a background was extracted from source-free regions (the outer 4${\tt '}$--6${\tt '}$ annulus)  for each XIS module separately. 
The spectral data  re-binning   was implemented  in order  to get  at least 20 cnts  per a spectral bin and to use  the $\chi^2$-statistic applying XSPEC v12.10.1 for our spectral fitting.  Thus, we made   the 0.3 -- 10 keV spectral fits for the XISs. it is worth noting that we did not use energies around 1.75 and 2.23 keV related to the XIS spectral problems around  Si and Au edges.


{\bf RXTE data}. We  analyzed the  data of 4U~1636--53 (1996 -- 2010) and SS~Cygni (1999 -- 2001, 2004 -- 2005 and 2010) obtained with {\it RXTE}~(see Bradt et al 1993). Observations made at different luminosity states of these sources were selected to study the characteristics of their outbursts.  In summary, we analyzed the RXTE sample of 4U~1636--53 data collected over 14 years for five intervals (marked with blue boxes in Fig.~\ref{variability_96-10} and listed in Table~\ref{tab:par_RXTE_1636}). Similarly, the RXTE sample for SS~Cygni accumulated over 11 years with five intervals shown in blue boxes in Fig.~\ref{variability_96-10_ss_cyg} and listed in Table \ref{tab:rxte_ss_cyg}) was analyzed. 
{\it RXTE}/PCA spectra have been extracted and analyzed, wherein  PCA {\it Standard 2} mode data, collected in the 3 -- 50~keV energy range, using the most recent release of PCA response calibration (ftool pcarmf v11.1). The relevant deadtime corrections to energy spectra 
have been applied. We used the data which  are available through the GSFC public archive\footnote{http://heasarc.gsfc.nasa.gov}. In Table~\ref{tab:par_RXTE_1636} we listed the groups  of {\it RXTE} observations of 4U~1636--53, which cover  the source evolution  from  faint 
 to   bright 
 phase events. 
Available   {\it RXTE} data contains three bright phase set ($R1-R3$)  and two  faint phase set ($R4$ -- $R5$). In  Table~\ref{tab:rxte_ss_cyg} we listed the groups  of {\it RXTE} observations of SS~Cygni, which cover  the source evolution  from  faint  to   bright  (phase) events. Thus, available   {\it RXTE} data includes four faint phase set ($Rs1$ -- $Rs4$) and one bright phase set ($Rs5$).  
The PCA energy spectra were modeled using XSPEC astrophysical fitting software. 
Systematic error of 0.5\% have been applied to the analyzed spectra. 

We have also used public 4U~1636--53 and SS~Cygni data from the  All-Sky Monitor (ASM) on-board \textit{RXTE}. The 4U~1636--53 light curve [1--12 keV ] shows long-term variability of mean soft flux during two $\sim$ six year intervals (Fig.~\ref{variability_96-10}). 
We  use definitions of the  fainter 
and brighter 
on luminosity phases 
to relate these phases 
to the source luminosity and we demonstrate that during the bright/faint phase 
transition of 4U~1636--53  the {\it CompTB} 
Normalization
changes from 
0.05 to 0.4 $L_{36}^{\rm soft}/{D^2}_{10}$
where  $L_{36}^{\rm soft}$ is the soft photon luminosity in units of $10^{36}$ erg/s  and  $D_{10}$ in units of  10 kpc  is distance to the source. 
As mentioned above, the {\it RXTE}/ASM light curve of 4U~1636--53  demonstrates long-term evolution of mean flux level: bright ($\sim$15 cts/s, 1997 -- 2000) and faint ($\sim$6 cts/s, 2001 -- 2010, see Fig.~\ref{variability_96-10}). This variability is consistent with  two states of $atoll$ sources: the $island$ (hard) and $banana$ (soft) states~\cite{hasinger89}. In turn, the light curve of SS~Cygni [1--12 keV] exhibits high X-ray variability at the level of 1--2 cnt/s with short-time bursts up to 7 cnt/s during 1996 -- 2009. For comparison, the object flared every $\sim$50 days from 12$^m$ to 8$^m$ in the optical range \cite{Lewin+vanderKlis06,Mauche+Robinson01}. In 2010, the object demonstrates an increase in the average level up to 3--4 cnt/s and a more prolonged burst with an amplitude of 7--8 cnt/s (Fig.~\ref{variability_96-10_ss_cyg}). 

%
%

\begin{figure}
\centering
\includegraphics[scale=.81,angle=0]{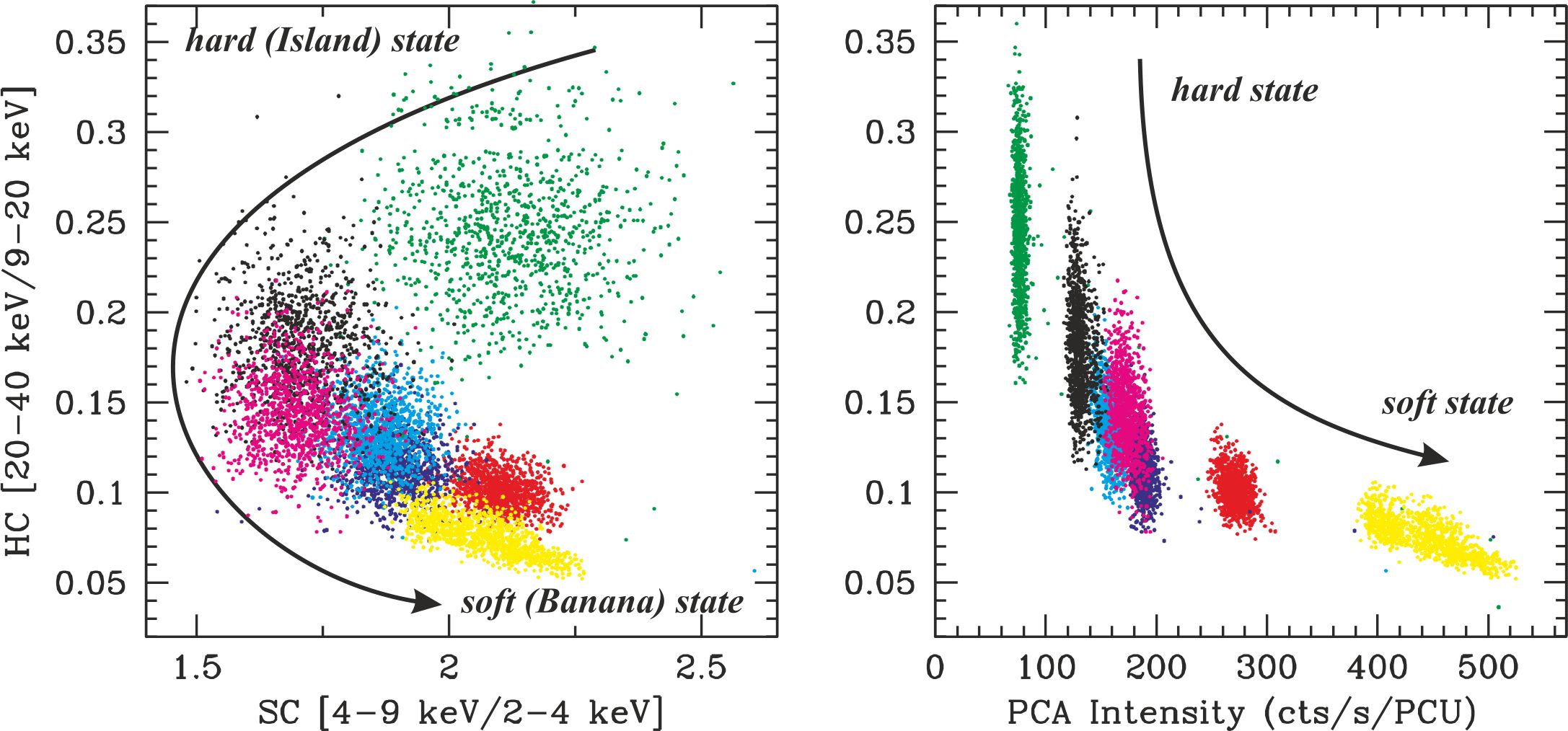}
\caption{
CCDs (left panel) and HIDs (right panel) of 4U~1636--53 for RXTE observations [ID=30053-02-01-001 ({\it red}), ID=60032-01-12-000 ({\it blue}), ID=60032-01-19-000 ({\it  bright blue}), ID=91027-01-01-000 ({\it green}), ID=93091-01-01-000 ({\it black}), ID=93091-01-02-000 ({\it crimson}), ID=10088-01-09-00 ({\it yellow})]  used in our analysis, with bin size 16 s. The typical error bars for the colors 
and intensity are negligible. The arrow shows the direction of development of the outburs from the hard state to the soft state.
}
\label{color_diagram_1636}
\end{figure}

%
%

\begin{figure}
\centering
\includegraphics[scale=.7,angle=0]{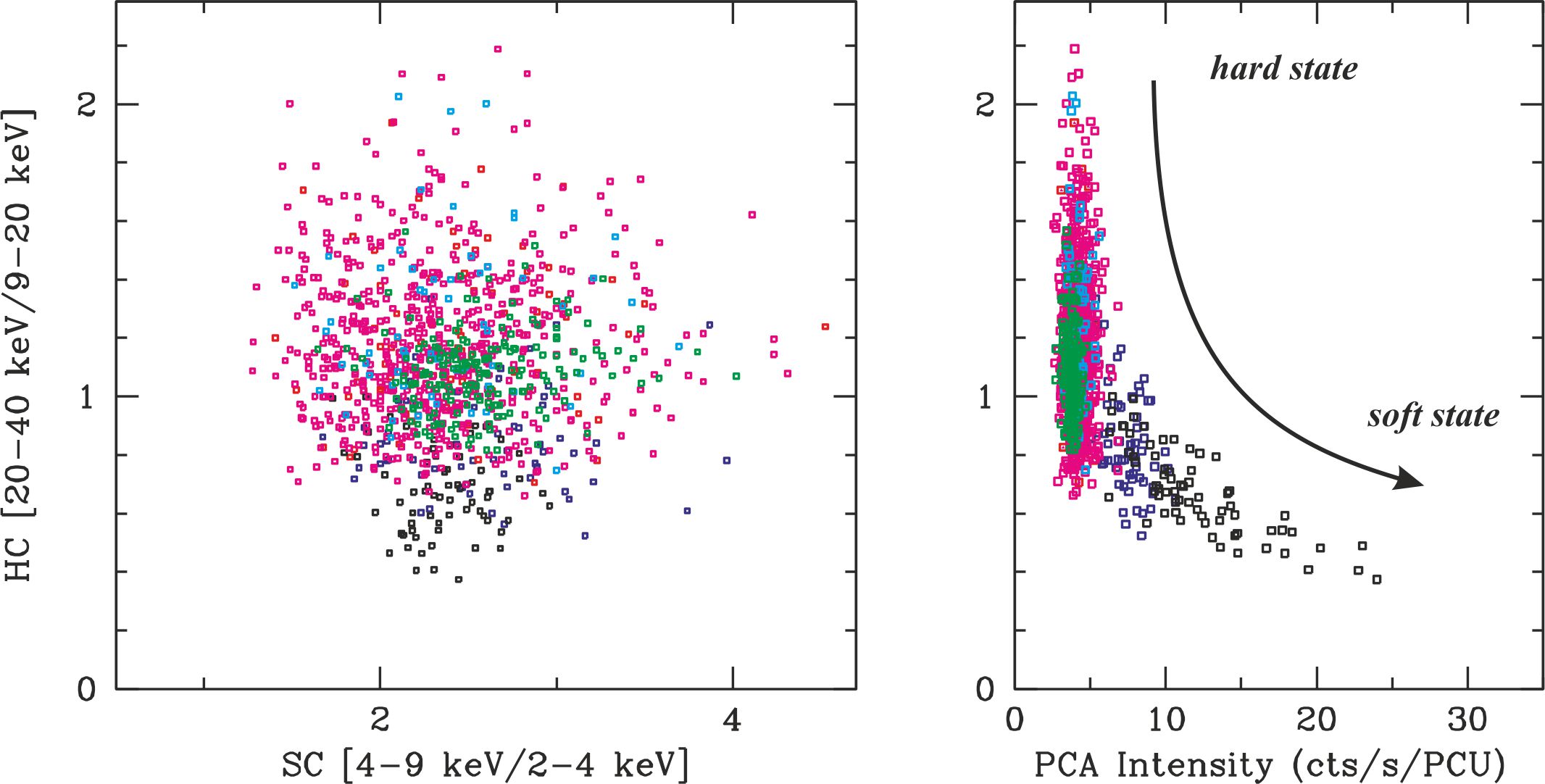}
\caption{CCDs (left panel) and HIDs (right panel) of SS~Cyg for RXTE observations [ ID=95421-01-01-05 ({\it crimson}), ID=95421-01-03-05 ({\it red}), ID=95421-01-01-01 ({\it blue}), ID=90007-01-01-00 ({\it black}), ID=50011-01-98-00 ({\it bright blue}), and ID=90007-01-19-00 ({\it green})]  used in our analysis, with bin size 16 s. The typical error bars for the colors and intensity are negligible. The arrow shows the direction of development of the outburst from the hard state to the soft state. 
}
\label{color_diagram_SS_Cyg}
\end{figure}

\section{RESULTS}

\subsection{Color-Color Diagrams \label{CCD}}

%
%

\begin{figure}
\centering
\includegraphics[scale=0.5, angle=0]{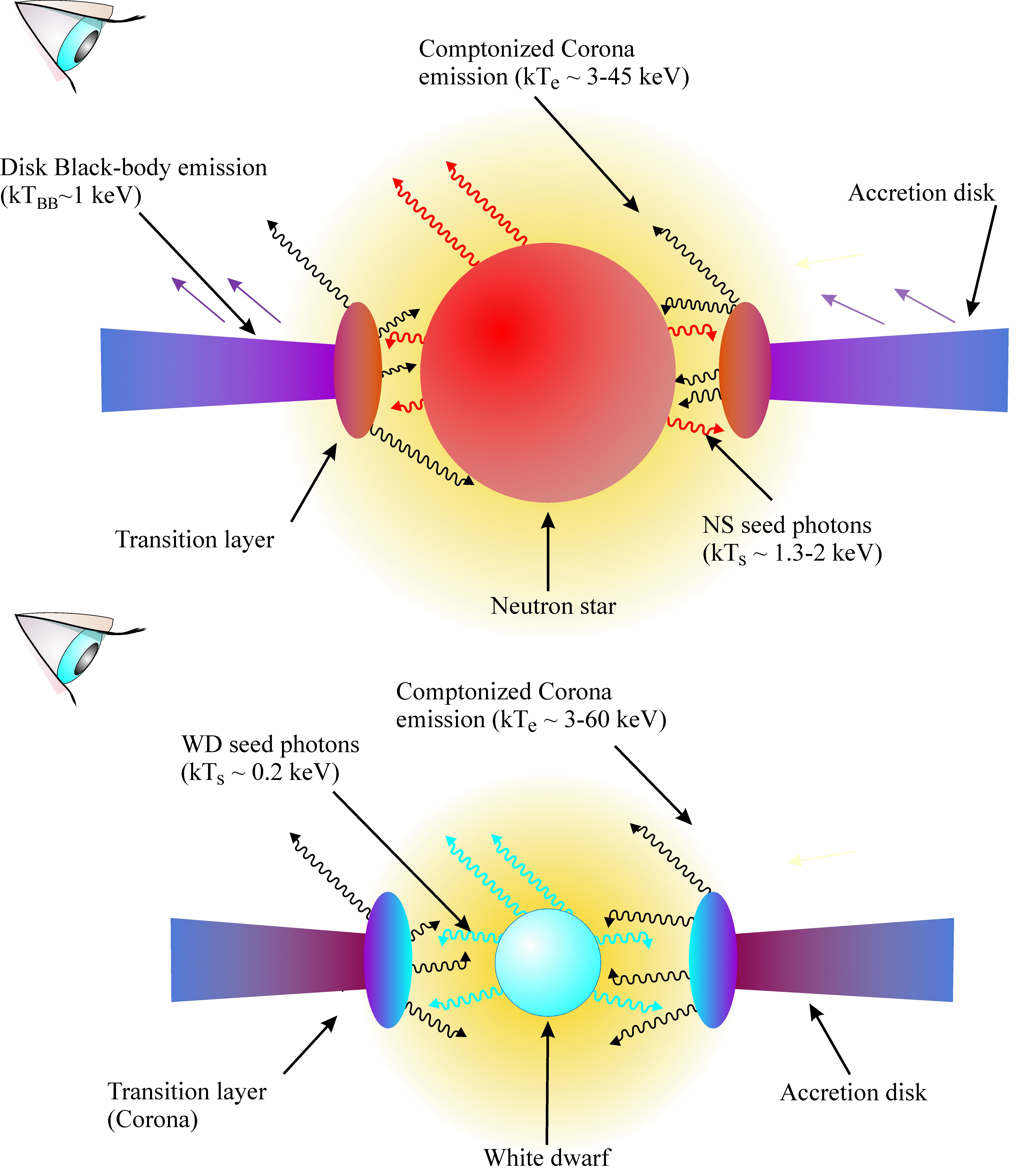}
\caption{A suggested  geometry of 4U~1636--53 (top) and SS~Cygni (bottom).   Disk and NS/WD soft photons are up-scattered off  hotter plasma of the TL located between the accretion disk and NS/WD surface.  Some fraction of these photons is seen directly by the Earth observer (blue arrows). Crimson and red photon trajectories correspond to TL and NS (upscattered) photons, respectively. NS surface reflects TL upscattered photons, while WD surface only absorbs TL upscattered photons.
}
\label{geometry}
\end{figure}

Before performing spectral analysis, we studied the properties of 4U~1636--53 and SS~Cygni in terms of color diagrams to trace the spectral state of the source as a first approximation. 
Namely, we investigate the light curves with 16s time binning of Standard-2 data for four energy channels 5 -- 10, 11 -- 24, 25 -- 54 and 55 -- 107. These channel ranges correspond to the 1.94 -- 4.05 keV, 4.05 -- 9.03 keV, 9.03 -- 20.3 keV and 20.3 -- 39.99 keV energy ranges, respectively. We constructed color-color diagrams (CCDs) of the sources using these energy-dependent light curves and defined the soft color (SC) as the ratio of count rates in the 4.05 -- 9.03 keV and 1.94 -- 4.05 keV energy bands, while the hard color (HC) is calculated as the ratio of count rates in the 20.3 -- 39.99 keV and 9.03 -- 20.3 keV energy ranges. 

The obtained CCDs and  hardness intensity diagrams (HIDs) are shown in Fig.~\ref{color_diagram_1636} for 4U~1636--53 and in Fig.~\ref{color_diagram_SS_Cyg} for SS~Cygni. 
From these figures we can see that the HID tracks of these two sources display a smooth and 
monotonic 
function, denoting to the similar, for these  two objects, physical mechanism of the hard/soft flux transition during a source evolution. In contrast to HIDs, CCD tracks for these two sources are fair different. 
So, for 4U~1636--53 the ranges of CS (1.5--2.5) and HC (0.05--0.35) are smaller than CS (1.2--5) and HC (0.4--2.2) for SS~Cygni. In addition, CCDs for 4U~1636--53 present a family of tracks, which are  elongated and characterized by a clear ``$C$''-like shape.   Although for SS~Cygni it resembles circular scattering and are characterized by a``$O$''-like shape. The typical error bars for the colors and  intensities 
are negligible.

\subsection{Spectral Analysis \label{spectral analysis}}

Initially, for both sources we have tried  a model consisting of an absorbed thermal Comptonization component ({\tt CompTB})  [the {\it CompTB} is a XSPEC contributed model\footnote{http://heasarc.gsfc.nasa.gov/docs/software/lheasoft---/xanadu/xspec/models/comptb.html}, \cite{F08}], 
but this model gave a poor description of data. Significant positive residuals around $\sim$~6.5 keV suggest 
a presence of the fluorescent iron emission line. 
An addition of a {\it Gaussian} line component at 6.4 keV 
considerably improves   the fitting quality of  both sources.   Therefore, we include  in the model  a simple {\it Gaussian} component,  which  parameters are  a centroid line energy $E_{line}$, the width of the line $\sigma_{line}$  and normalization, $N_{line}$ to fit the data in the 6 -- 8 keV  energy range. 
Furthermore, an addition of a thermal component ({\tt bbody}) to the spectra of 4U~1636--53 further improves fit quality and provides a statistically acceptable $\chi^2_{red}$. At the first time the fluorescent iron emission line in 4U~1636--53  was detected by \cite{Lyu14} using  {\it Suzaku}, {\it XMM}-Newton, and {\it RXTE}. 
\cite{Lyu14} successfully described this emission feature with the  {\it Gaussian} line model. They used a model consisting of thermal components [representing the accretion disk by {\tt Diskbb}  \cite{Mitsuda84,Makishima86},  the NS surface and boundary layer by {\tt Bbody}; for both of the models  dominating energy release  are around 1 keV], and a Comptonized component [representing a hot corona by {\tt Nthcomp} \cite{Zdziarski96,Zycki99}], and either a {\tt Gaussian} or a relativistic line component of the model of  an iron emission line at $\sim$ 6.5 keV. 

 We also use   the interstellar absorption with a column density $N_H$ in the model. It should be noted  that we  fixed certain parameters of the {\it CompTB} component: 
$\gamma=3$ (low energy index of the seed photon spectrum) and $\delta=0$ because we neglect an efficiency  of the  bulk inflow effect versus the  thermal Comptonization   for  NS  4U~1636--53 and WD SS~Cygni. We apply a value of hydrogen column $N_H=3.2\times 10^{21}$ cm$^{-2}$ (for 4U~1636--53, \cite{fiocchi06}) and  $N_H=0.2\times 10^{22}$ cm$^{-2}$ (for SS~Cygni, \cite{Madejski99}). 
As a result, in our spectral  data analysis for 4U~1636--53 we use a model which consists a  sum of a  Comptonization   component ({\it CompTB}), 
a soft blackbody component of the temperature,  $T_{BB}$ and the {\it Gaussian} line component.  We also modeled the SS~Cygni spectra using a combination of {\it CompTB} component  and iron line components. 
The {\it CompTB} spectral component has the following parameters:   the seed photon  temperature,  $T_s$, the energy index of the Comptonization spectrum $\alpha$ ($=\Gamma-1$), the electron temperature $T_e$,   a Comptonization  fraction $f$ [$f=A/(1+A)$, which is the relative weight of the Comptonization component  and normalization of the seed photon spectrum, $N_{com}$. 

%
%
\begin{figure}
\centering
 \includegraphics[scale=0.8,angle=0]{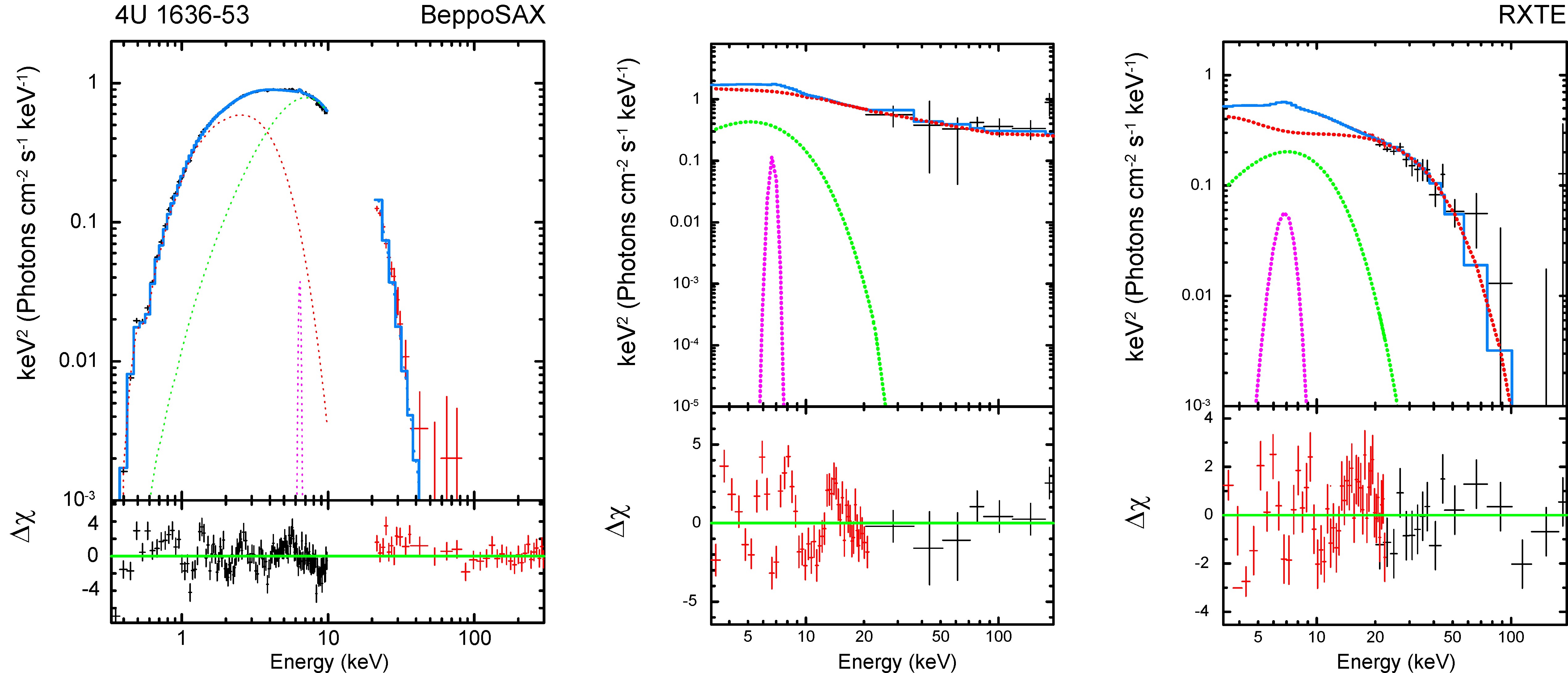}
\caption{
Left: the best-fit spectrum of 4U~1636--53 
 in $E*F(E)$ units using {\it Beppo}SAX observation  20836001 carried out on 15 -- 17 February 2000.   
%
Examples of typical $E*F(E)$ spectral diagram  of 4U~1636--53 during hard (central panel) and soft (right panel) state events with RXTE  in the model {\it tbabs * (Bbody + CompTB + Gaussian)}  
for high-luminosity (banana) state [10088-01-09-02 observation] 
and low-luminosity (island) state [94437-01-01-010 observation]. 
}
\label{BeppoSAX_spectra}
\end{figure}

In Figure~\ref{geometry}  we illustrate  our spectral model as a basic model for  fitting  the 
X-ray spectral data for 4U~1636--53 (top, NS case) and SS~Cygni (bottom, WD case). We assume that accretion onto a NS/WD 
takes place when the material passing through the two main regions:  a geometrically thin accretion disk [the standard Shakura-Sunyaev disk, see \cite{ss73}] and the 
TL, where a NS/WD and  disk soft photons are  upscattered off hot electrons (see, Titarchuk et al.1998). 
In our picture, the emergent thermal Comptonization spectrum is  formed in the  TL, where thermal disk  seed photons and soft photons from the NS/WD  are up-scattered off the relatively hot plasma. 
Some fraction of these seed soft photons can be also seen directly (blue arrows for a NS case). Recall that the accretion disk around WD (or CVs) is usually too cold ($kT\ll 1$ keV) and does not emit X-rays. 
Red and black photon trajectories shown in Fig. \ref{geometry}  correspond to soft (seed) and hard (up-scattered) photons, respectively. Furthermore, the NS surface is capable of scattering/reflecting the  TL photons. In contrast to a NS, the WD surface in nonmagnetic CVs will no scatter/reflect TL photons but  observed them. It is  colder than that in a NS case.

%
%

\subsubsection{4U~1636--53 \label{spectral analysis_1636}}

We show examples of X-ray spectra of 4U~1636--53  in  Fig.~\ref{BeppoSAX_spectra} 
using $Beppo$SAX (left panel) and {\it RXTE} (central and right panels) data, respectively.
Spectral analysis of {\it Beppo}SAX and  {\it RXTE}  observations indicates  that X-ray  spectra of 4U~1636--53 can be   described by a model with a  Comptonization component  represented by  the {\it CompTB} model. 

%
%
\begin{table*}
  \centering 
 \caption{Best-fit parameters of spectral analysis of {\it Beppo}SAX observations of 4U~1636--53 in 0.3--50~keV energy range$^{\dagger}$.
Parameter errors correspond to 1$\sigma$ confidence level.}
 \begin{tabular}{cccccccccccccc}
      \hline\hline
Set & $T_{bb}$, & $N_{bb}^{\dagger\dagger}$ &$T_s$, & $\alpha=$  & $T_e,$ & $\log(A)$ & N$_{com}^{\dagger\dagger}$ &  E$_{line}$,&   $N_{line}^{\dagger\dagger}$ &  $\chi^2_{red}$ \\
             &  keV         &                           & keV  &$\Gamma-1$  & keV    &                                  &                                   &   keV       &                               &                       (dof) \\
\hline
B1  &  0.71(8) &  39.7(2) &  1.99(1)  & 1.00(4) & 2.36(8) & 0.04(1) & 5.35(1) & 6.4(2)  &0.14(1) & 1.08(219)\\
B2  &   0.70(6) &  26.7(1) & 1.97(1) & 1.00(4) & 2.54(9) & 0.66(2) & 12.59(1) & 6.5(2)  &0.06(1) & 1.09(168)\\
B3  &   0.72(9) &  20.5(1) & 1.80(1) & 0.98(3) & 3.21(7) & 0.48(7) & 5.81(3) & 6.4(1)  &0.02(1) & 0.83(363)\\
\hline
      \end{tabular}
    \label{tab:fit_table_sax_1636}
\\
$^\dagger$ The spectral model is  {\it tbabs*(Blackbody + CompTB + Gaussian)}, normalization parameters of {\it blackbdody} and {\it CompTB} components are in units of $L_{35}^{\rm soft}/d^2_{10}$, 
where $L_{35}^{\rm soft}$ is 
the soft photon  luminosity in units of 10$^{35}$; 
$d_{10}$ is the distance to the source in units of 10 kpc 
and {\it Gaussian} component is in units of $10^{-2}\times$ total photons cm$^{-2}$s$^{-1}$ in line.
\end{table*}

On the top of Figure~\ref{BeppoSAX_spectra} we demonstrate  the best-fit {\it Beppo}SAX spectrum of 4U~1636--53 using  our   model for  {\it Beppo}SAX observation (id=20836001) carried out on 15--17 February, 2000.  The data are presented by crosses and the best-fit spectral  model,   {\tt tbabs * (Blackbody + CompTB + Gaussian)} by green line. The model components  are shown by blue, red and crimson lines for  the {\tt blackbdody}, {\tt CompTB}  and {\tt Gaussian} components, respectively. 
On the bottom we show $\Delta \chi$ vs    photon energy in keV. From Fig.~\ref{BeppoSAX_spectra} we can see that  the  low-energy end of {\it Beppo}SAX spectrum the 4U~1636--53 the source emission is  modified  by  photoelectric absorption. 
The best-fit model parameters are $\Gamma$=1.98$\pm$0.03, $T_e$=3.68$\pm$0.05 keV and $E_{line}$=6.4$\pm$0.1 keV (reduced $\chi^2$=0.83 for 363 d.o.f) (see more details in Table~\ref{tab:fit_table_sax_1636}). In particular, we  find that  an addition of  the soft thermal   
component of the   temperature $kT_{BB}=$0.6$-$0.7 keV   significantly improves  the fit quality of the  {\it Beppo}SAX  spectra. For the {\it Beppo}SAX data 
(see Tables~\ref{tab:fit_table_sax_1636}) we find that the spectral index $\alpha$ is 
of 1.03$\pm$0.04 (or the corresponding photon index $\Gamma=\alpha+1$ is 
 2.03$\pm$0.04).  The color temperature $kT_{s}$  of the {\it CompTB} component changes from  1.2 keV to  1.7 keV,  which is consistent with that using  the {\it Beppo}SAX data set of our analysis (see Table~\ref{tab:fit_table_sax_1636}) and previous NS studies by  \cite{balman01} 
and  \cite{ch06}. We should also emphasize  that the  temperature of the seed photons $kT_s$  of the {\it CompTB} component usually  increases up to 1.7 keV in the fainter phases  and generally decreases to 1.2 keV in the bright phases. 

%
%
\begin{table*}
  \centering 
 \caption{Best-fit parameters of spectral analysis of $ASCA$ observations of 4U~1636--53 in 0.6--10~keV energy range$^{\dagger}$.
Parameter errors correspond to 1$\sigma$ confidence level.}
 \begin{tabular}{cccccccccccccc}
      \hline\hline
Set &  $T_{bb}$, & $N_{bb}^{\dagger\dagger}$ &$T_s$, & $\alpha=$  & $T_e,$ & $\log(A)$ & N$_{com}^{\dagger\dagger}$ &  E$_{line}$,&   $N_{line}^{\dagger\dagger}$ &  $\chi^2_{red}$ \\
            & keV         &                           & keV  &$\Gamma-1$  & keV    &                                  &                                   &   keV       &                               &                       (dof) \\
\hline
\\
A1 &   0.70(3) & 25.3(3)  & 1.81(2) & 1.00(4) & 3.0(2) & 1.00$^{\dagger\dagger\dagger}$ & 5.14(2) & 6.4(3) &  0.21(1) & 1.08(647) \\
A2 &  0.75(7) & 22.8(1)  & 1.83(1) & 1.01(5) & 2.5(1) & 0.42(4) & 5.54(1) & 6.4(2) &  0.14(3) & 1.13(769)\\
A3 &   0.69(8) & 26.8(1)  & 1.81(1) & 1.00(3) & 2.4(1) & 0.43(1) & 6.18(2) & 6.5(2) &  0.01(1) & 1.17(794)\\
A4 &  0.71(4) & 29.8(3)  & 1.79(2)  & 0.96(4) & 2.3(1) & 0.43(2) & 6.62(4) & 6.4(1) &  0.02(1) & 1.14(638)\\
A5 &   0.70(5) & 35.8(2)  & 1.84(1) & 1.00(5) & 2.4(1) & 0.38(2) & 6.14(2) & 6.5(2) &  0.05(1) & 1.15(801)\\
A6 &   0.75(8) & 20.9(1)  & 1.81(1)  & 1.02(4) & 2.5(1) & 0.37(2) & 3.91(1) & 6.4(1) &  0.01(1) & 1.10(790)\\
A7 &   0.72(4) & 23.2(1)  & 1.80(1) & 1.00(3) & 2.6(1) & 0.41(3) & 5.56(3) & 6.5(2) &  0.04(1) & 1.00(647)\\
A8 &   0.71(8) & 16.5(1)  &  1.90(2) & 0.99(5) & 2.7(1) & 0.54(1) & 4.71(1) & 6.4(1) &  0.01(1) & 1.02(739)\\
A9 &    0.70(3) & 14.3(1)  & 1.74(1) & 1.01(4) & 2.5(8) & 0.35(1) & 3.34(1) & 6.4(2) &  0.01(1) & 1.11(771)\\
\hline
      \end{tabular}
    \label{tab:fit_table_asca_1636}
\\$^\dagger$ The spectral model is  {\it tbabs*(Blackbody + CompTB + Gaussian)},
normalization parameters of {\it Blackbody} and {\it CompTB} components are in units of 
$L_{35}^{\rm soft}/d^2_{10}$, 
where $L_{35}^{\rm soft}$ is 
the soft photon  luminosity in units of 10$^{35}$; 
$d_{10}$ is the distance to the source in units of 10 kpc 
and {\it Gaussian} component is in units of $10^{-2}\times$ total photons cm$^{-2}$s$^{-1}$ in line; 
$^{\dagger\dagger\dagger}$ when parameter $\log(A)\gg1$, it is fixed to a value 1.0 (see comments in the text). 
\end{table*}

However, the main part of ASCA observations (A1--A8) fells on the initial observation interval, up to 1996 (MJD $\sim$50000), when ASM/{\it RXTE} did not perform X-ray monitoring. The spectral analysis for these ASCA observations of 4U~1636--53 indicates a typical active state of the source with an increased X-ray luminosity in average. During these observations, the plasma temperature turns out to be low $T_e\sim$ 2--4 keV, the seed photon temperature varied slightly at the level of 1.2--1.7 keV, but the spectral index showed quasi-constancy at the $\alpha\sim$ 1 level (see Table 
\ref{tab:fit_table_asca_1636}).

%
%
\begin{figure*}
\centering
\includegraphics[scale=0.6,angle=0]{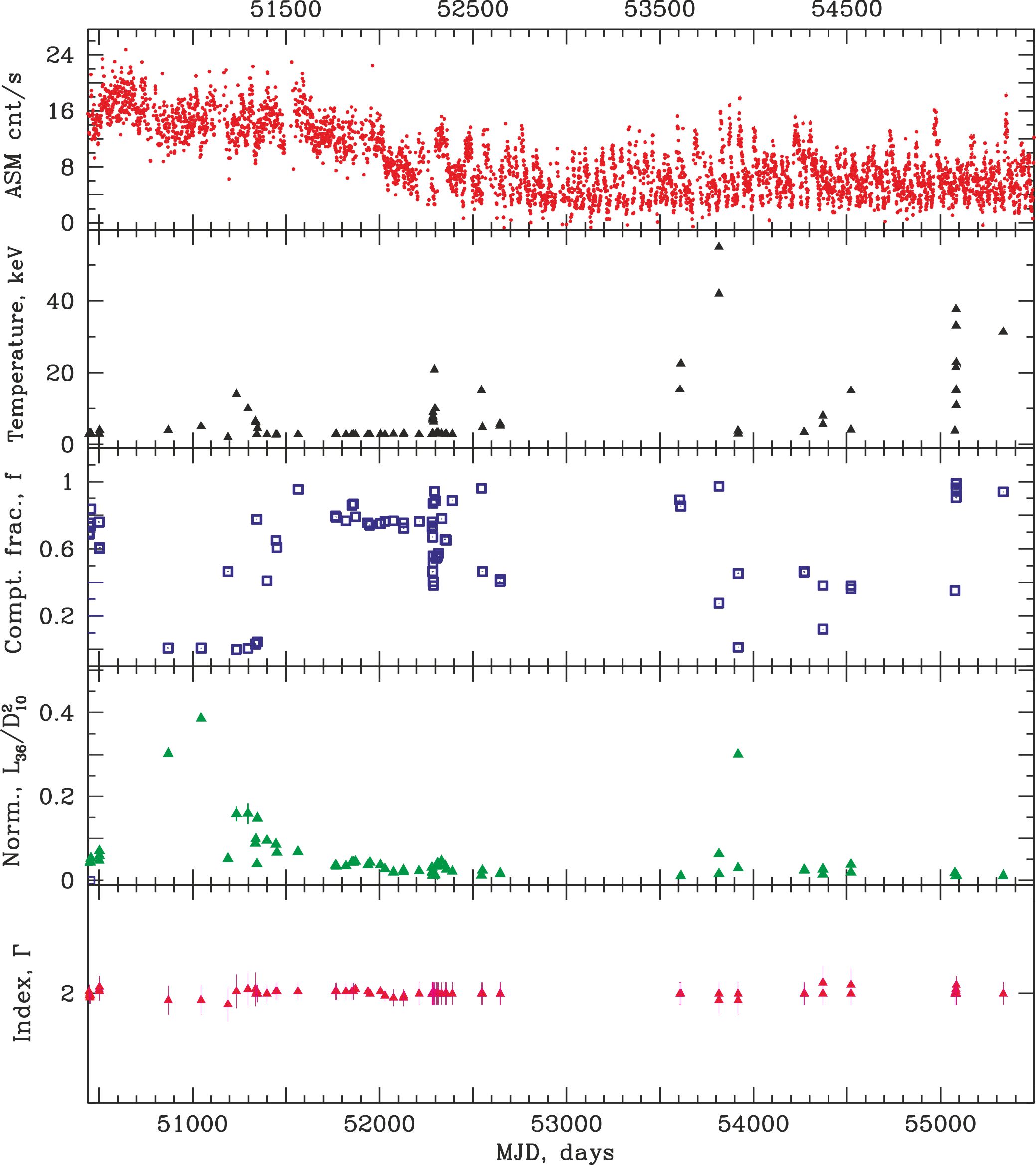}
\caption{{\it From Top to Bottom:}
Evolutions of the {\it RXTE}/ASM count rate (red),
electron temperature $kT_e$ (black),   
Comptonization  fraction $f$ (blue), {\it CompTB} normalization ({\it blue})  and  photon index $\Gamma$ ({\it crimson}) during 1996 -- 2010 outburst transitions 
for NS binary 4U~1636--53. 
}
\label{lc_1998}
\end{figure*}

Furthermore, the spectral model {\it tbabs * (Bbody + CompTB + Gaussian)}, tested on {\it Beppo}SAX and ASCA observations, was applied to the {\it RXTE} data of 4U1636--53. Unfortunately  {\it RXTE} detectors cannot provide well calibrated spectra  below 3 keV  while the  broad energy band of {\it Beppo}SAX telescopes allows us to determine  the parameters of {\it Blackbody} components  at low energies.  Thus in order to fit the {\it RXTE} data  we have to fix the  temperature of {\it blackbdody} component at a value of $kT_{BB}=$ 0.6 keV obtained as an upper limit  in  our   analyze of   the {\it Beppo}SAX data. The  best-fit spectral parameters using {\it RXTE} observations are presented in Fig.~\ref{lc_1998}. In particular,  we find that electron temperature $kT_e$ of the  {\it CompTB} component varies from 2.3 to 48 keV,  while the photon   index $\Gamma$  is almost constant ($\Gamma=1.99\pm 0.02$)  for all observations.  It is worth noting that the width $\sigma_{line}$ of {\it Gaussian} component does not  vary much  and it is  in the range of 0.5 -- 0.8 keV.

In  Figure~\ref{BeppoSAX_spectra} 
we  show examples of the typical photon spectra [$E*F(E)$ spectral diagrams] of 4U~1636--53 during the fainter phase (94437-01-01-010,  right panel) and the brighter phases 10088-01-09-02,  central panel) detected by {\it RXTE} on MJD 55084.6 and 50503.2, respectively. The adopted spectral model shows a very good fidelity throughout all data sets used in our analysis.  Red, green and  purple lines 
stand for the {\it CompTB, Blackbody},  and  {\it Gaussian} components, respectively. While the resulting spectrum is shown by a  blue line. In the bottom panels we indicate residuals. The best-fit model parameters are (for soft state) $\Gamma$=1.99$\pm$0.02, $T_e$=2.94$\pm$0.01 keV and $E_{line}$=6.53$\pm$0.06 keV;  and  (for hard state) $\Gamma$=2.00$\pm$0.04, $T_e$=12.54$\pm$0.09 keV and $E_{gauss}$=6.35$\pm$0.04 keV, respectively (Fig.~\ref{lc_1998}).
A value of reduced $\chi^2_{red}=\chi^2/N_{dof}$, where $N_{dof}$ is a number of degree of freedom, is  less or about 1.0 for most observations.

We find  some differences between our values of the best-fit model parameters and those in the literature.  In particular, the photon index  $\Gamma$, estimated by \cite{Lyu14}  for observation  IDs=91027-01-01-000, 93091-01-01-000, 93091-01-02-000, 94310-01-02-03, 94310-01-02-04, 94310-01-02-05, 94310-01-02-02, 94310-01-03-000, 94310-01-03-00, and  94310-01-04-00, varies from 1.8 to 2.7, while we got a more modest variation from 1.8 to 2.2. 
This discrepancy in the index $\Gamma$ may be the result of using somewhat different spectral models by us and by \cite{Lyu14}.  They used the {\it wabs*(Diskbb + Bbody + NthComp + Gaussian)} spectral model and applied it to a slightly different set of observations by {\it RXTE}. It is also possible that such a discrepancy is associated with a slightly different energy range for fitting the combined {\it RXTE} and XMM-{\it Newton} spectra used by \cite{Lyu14}.

Such comparisons indicate that it is important to take into account the contribution of radiation from a NS  at low energies in the spectrum. Using broadband observations of {\it Beppo}SAX and ASCA, we can well determine all the parameters of our spectral model, while thanks to extensive observations of 4U~1636--53 with {\it RXTE}, we can explore the long-term general behavior of the source during the  spectral transitions in the  3 -- 50 keV energy range.

 \subsubsection{SS~Cygni\label{spectral analysis_SS_Cyg}}

{\it Suzaku} observations of SS~Cygni  made it possible to study the source at  the quiescence ($S1$ and $S3$, see Table \ref{tab:table_suzaku_ss_cyg}) and in the outburst ($S2$). We have shown that the X-ray source continuum can be described by the Comptonization model with the maximum plasma temperature at quiescence $\sim$20 keV, which is much higher than in the outburst $\sim$6 keV. The degree of illumination $f$ is $\sim$1 at the quiescence. 
We consider the standard optically thin TL as the most plausible picture of the plasma configuration at the quiescence. The degree of illumination in the outburst state $f= 0.2$ and the broad line of iron 6.4 keV indicate that the reflection in the outburst state comes from the accretion disk and the equatorial accretion ``belt''. The broad line at 6.4 keV indicates that the optically thin thermal plasma is distributed over the accretion disk like a corona. 

In Figure \ref{suzaku_hard_soft_state_spectrum_ss_cyg}, we present the spectral components of SS~Cygni in the 0.3--20 keV band: the Comptonized thermal plasma model provides a description of the hard X-ray emission ({\it red}), and the reflection of the hard X-ray emission by an accretion disk and/or the surface  while the disk gives the observed iron fluorescence lines K$_{\alpha}$ and hump in a continuum ({\it green}; \cite{Done_Osborne97,Ishida09}). It is clearly seen that the energy range of $\sim$6.4--7.0 keV also contains other lines associated with neutral, He-like, and H-like K$_{\alpha}$ Fe lines. We found a number of  positive (emission) excesses in the spectrum, and in order to better describe the spectrum, we added a number of additional lines. In this approach, we found lines at energies of 1.02, 1.85, 2.8, 6.39, 6.67, and 6.97 keV, which are easily associated with  Ne~X, Si~XIII, S~XV, Fe~I--XXII, Fe~XXV, and Fe~XXVI K$_{\alpha}$ lines, respectively.

%
%
\begin{figure*}
\centering
\includegraphics[scale=0.82,angle=0]{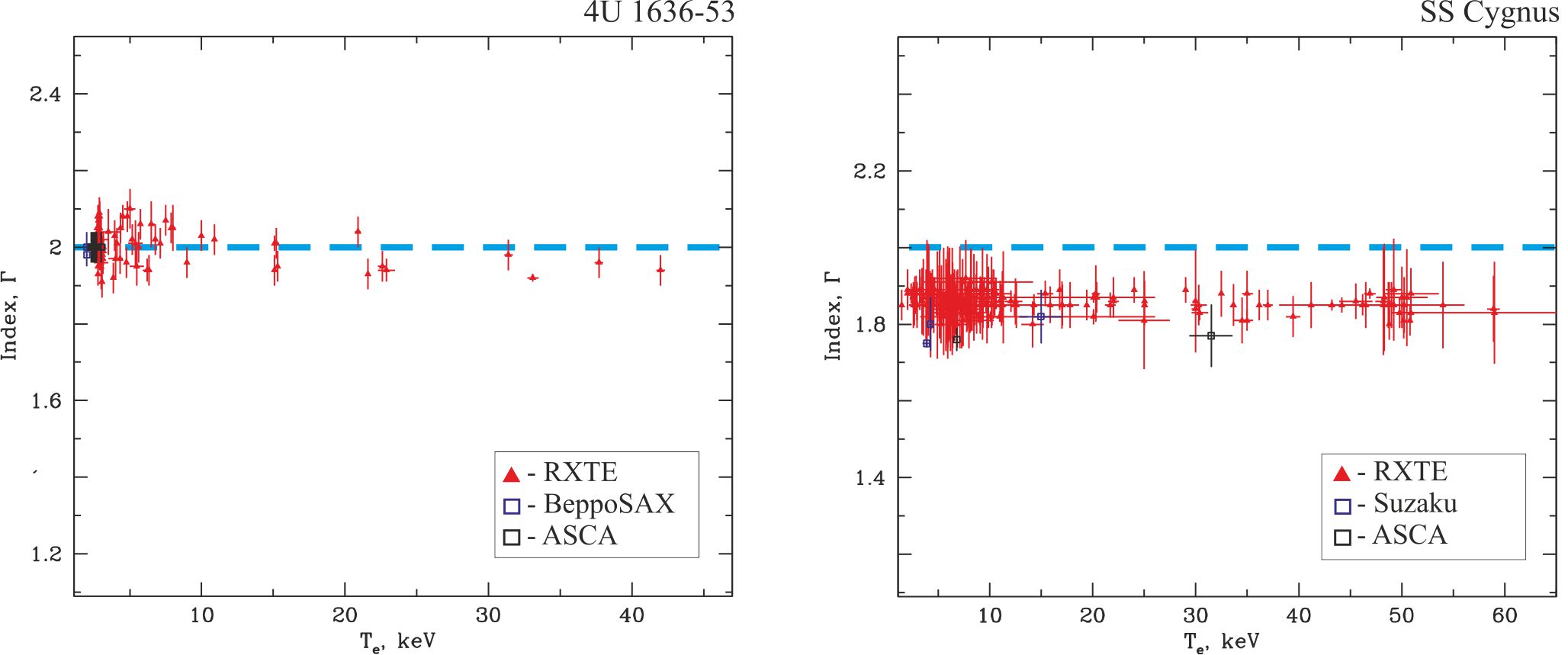}
\caption{
The photon index plotted versus the  electron temperature $T_e$ (in keV) for 4U~1636--53 (left) and SS~Cyg (right). 
}
\label{gam_Te}
\end{figure*}

We also analyze archived ASCA X-ray data from SS~Cygni both at rest and in outburst using Comptonization model for continuum and {\it Gaussian} profiles for line emission. 
The best-fit model parameters of the model are given in Table~\ref{tab:fit_table_suzaku+asca_ss_cyg}. The difference between the obtained  TL and WD surface temperatures indicates that the only reflector is the surface of the inner part (TL) of the disk, and not the white dwarf. The degree of illumination of TL $f\sim$ 1 in the outburst also more closely matches the hard X-ray emission that forms the corona above the surface of the WD, and not just the equatorial ``belt''. We detect a partially ionized absorption in the ASCA outburst spectrum, which is likely the X-ray signature of the outflowing wind.  The ASCA data also allow a detailed study of SS~Cygni at the quiescence when  the source spectrum becomes harder. The outburst spectrum, on the other hand, is much softer. Although the outburst spectrum of SS~Cygni is much softer than the quiescence spectrum, the total luminosity of the X-ray component is the same in both cases, indicating the importance of optically thin plasma emission even it the outburst. 
We note that the {\it Suzaku} and {\it ASCA} observations of SS~Cygni make it possible to resolve many lines and thus, take into account the complex pattern of reflection in SS~Cygni. 

Figure~\ref{
rxte_hard_soft_state_spectrum_ss_cyg} 
presents examples of the typical  $E*F(E)$ spectral diagram  of SS~Cygni during the low-soft state (LSS, {\it left} panel), the low-hard state (LHS, {\it central left} panel), the intermediate state (IS, {\it central right} panel)  and the high-soft state (HSS, {\it right} panel) events. The best-fit {\it RXTE} spectra ({\it top} panels) in the model $tbabs*(CompTB+Gaussian)$  with $\Delta\chi$ ({\it bottom} panels) for the LSS [95421-01-01-05 observation,  $\chi^2_{red}$=0.93 for 54 dof], LHS [95421-01-03-06,  $\chi^2_{red}$=0.94 for 54 dof], IS [90007-01-05-00,  $\chi^2_{red}$=0.98 for 54 dof],  and HSS [40012-01-15-00,  $\chi^2_{red}$=0.86 for 54 dof].
The best-fit model parameters are 
$\Gamma$=1.85$\pm$0.08, $T_e$=2.40$\pm$0.06 keV and $E_{line}$=6.6$\pm$0.2 keV (for LSS) ;  
$\Gamma$=1.85$\pm$0.07, $T_e$=11.0$\pm$0.4 keV and $E_{line}$=6.6$\pm$0.1 keV (for LHS);  
$\Gamma$=1.85$\pm$0.09, $T_e$=7.6$\pm$0.6 keV and $E_{line}$=6.64$\pm$0.3 keV (for IS);  
and  
$\Gamma$=1.8$\pm$0.1, $T_e$=7.9$\pm$0.2 keV and $E_{line}$=6.67$\pm$0.06 keV (for HSS), respectively.   
{\it Green} and {\it violet} lines stand {\it CompTB} and {\it Gaussian} components, respectively.

%
%
\begin{figure*}
\centering
\includegraphics[scale=0.7,angle=0]{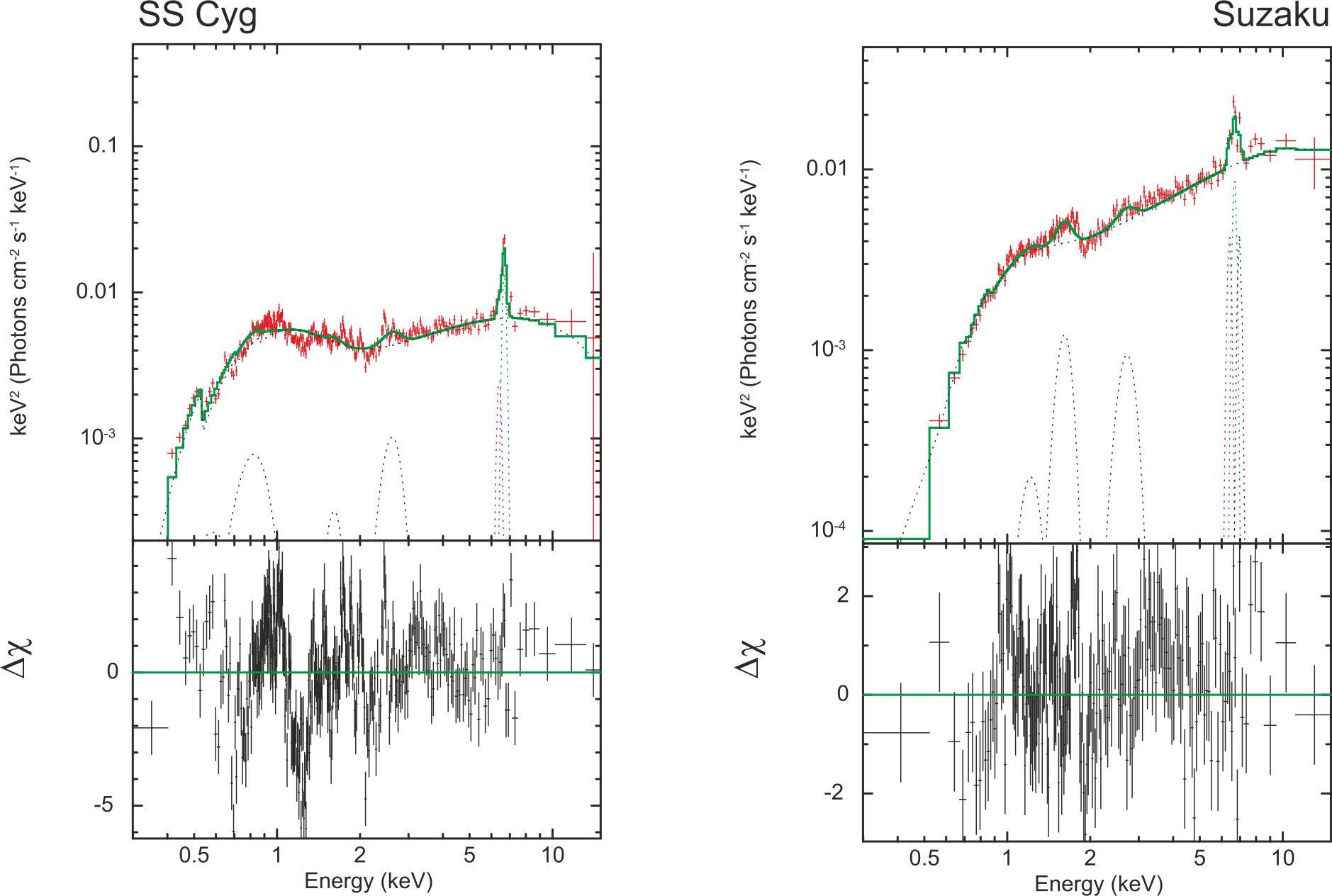}
\caption{Examples of typical $E*F(E)$ spectral diagram  of SS~Cyg during the high-soft state (HSS, {\it left} panel) and   the low-hard state (LHS, {\it right} panel) events. The best-fit Suzaku spectra ({\it top} panels) in the model {\it tbabs*(CompTB + Gaussian)}  with $\Delta\chi$ ({\it bottom} panels) for the LHS [400007010 observation,  $\chi^2_{red}$=0.93 for 54 dof] and the HSS [109015010,  $\chi^2_{red}$=0.86 for 54 dof]. Green and violet lines stand show  the {\it CompTB} and {\it Gaussian} components, respectively.
}
\label{suzaku_hard_soft_state_spectrum_ss_cyg}
\end{figure*}

%
%
\begin{figure*}
\centering
\includegraphics[scale=0.88,angle=0]{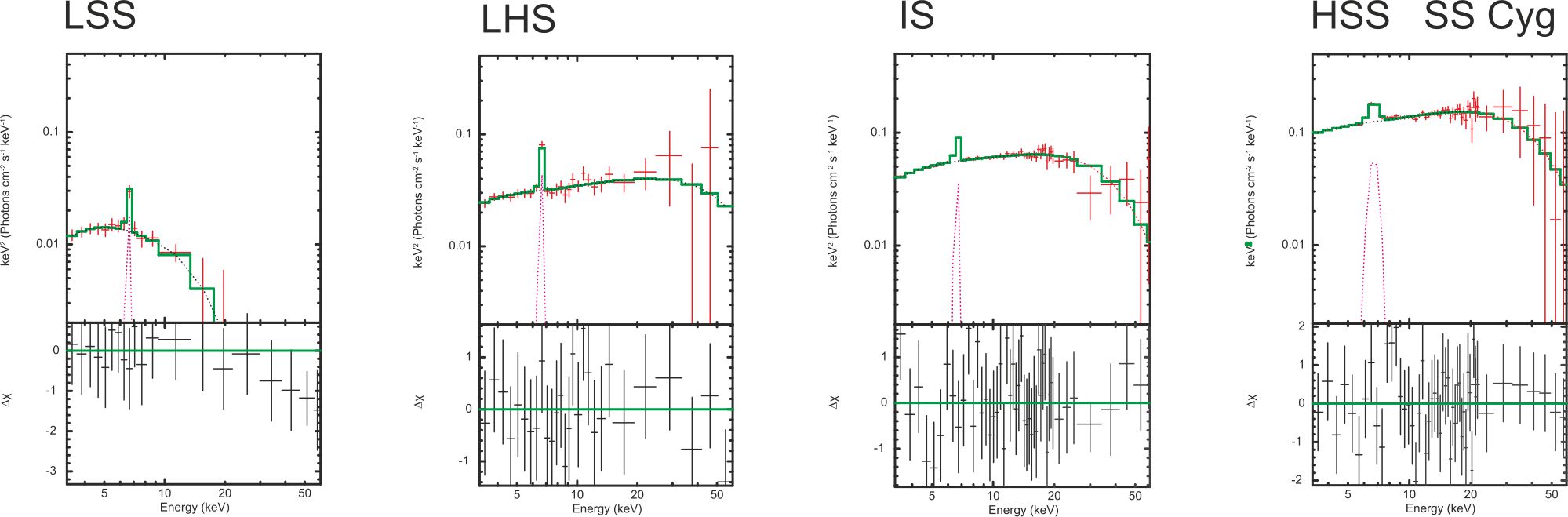}
\caption{Examples of typical $E*F(E)$ spectral diagram  of SS~Cyg during the low-soft state
 (LSS, left panel), 
the low-hard state (LHS, {\it central left} panel), the  intermediate state (IS, {\it central right} panel)  and the high-soft state (HSS, {\it right} panel) events. The best-fit {\it RXTE} spectra ({\it top} panels) in the model {\it tbabs * (CompTB + Gaussian)}  with $\Delta\chi$ ({\it bottom} panels) for the LSS [95421-01-01-05 observation,  $\chi^2_{red}$=0.93 for 54 dof], the LHS [95421-01-03-06,  $\chi^2_{red}$=0.94 for 54 dof], the IS [90007-01-05-00,  $\chi^2_{red}$=0.98 for 54 dof],  and the HSS [40012-01-15-00,  $\chi^2_{red}$=0.86 for 54 dof]. {\it Green} and {\it violet} lines stand {\it Bbody}, {\it CompTB} and {\it Gaussian} components, respectively.
}
\label{rxte_hard_soft_state_spectrum_ss_cyg}
\end{figure*}

%
%
\begin{table*}
  \centering 
 \caption{Best-fit parameters of spectral analysis of {\it Suzaku} and {\it ASCA} observations of SS~Cygni in 0.3--10~keV energy range$^{\dagger}$.
Parameter errors correspond to 1$\sigma$ confidence level.}
 \begin{tabular}{lllllllllcccccccccc}
      \hline\hline
ObsID&         &  S1   & S2 & S3   & As1  & As2    & Line transition\\
Model                  & Parameter    &                   &                   &                   &                   &                 &  identification\\

\hline
{\it CompTB} & $T_s$, keV &  0.20(7) & 0.23(1) & 0.20(7) & 0.21(7)  & 0.23(2) & \\
             & $\alpha$    &   0.75(9)  &  0.80(2)  & 0.8(1)  & 0.76(5)  & 0.77(2)  &\\
             &  $\log(A)$  &  0.13(5)               &  0.22(4)                &   0.5(1)                &    0.2(1)             &   0.78(3)            & \\
             & $T_e$, keV &    15(2)             &    3.9(4)              &  4.3(1)                 &  32(1)              &  6.8(1)            & \\
             & N$_{com}^{\dagger\dagger}$ &   4.3(6)               & 3.7(1)                 &   1.6(2)                &   3.19(7)             & 1.24(8)             & \\
{\it Gauss}$_1$ &E$_{l1}$, keV &     6.94(8)             &    6.95(4)              &   6.91(8)                &   6.9(1)             & 6.95(9)             & Fe~XXVII $K_{\alpha}$ \\
             &  $N_{1}^{\dagger\dagger}$ &    0.31(7)             &    0.32(2)              &   0.3(6)                &   0.3(1)             &  0.3(1)           &  \\
{\it Gauss}$_2$ &E$_{l2}$, keV &     6.69(4)             &    6.67(6)              &   6.68(5)                &   6.67(9)             & 6.67(2)             & Fe~XXVI $K_{\alpha}$ \\
             &  $N_{2}^{\dagger\dagger}$ &    0.22(6)             &    0.23(4)              &   0.21(7)                &   0.2(1)             &  0.20(5)        &     \\
{\it Gauss}$_3$ &E$_{l3}$, keV &     6.39(7)             &    6.41(9)              &   6.40(5)                &   6.46(8)             & 6.4(2)              & Fe~I $K_{\alpha}$ \\
             &  $N_{l3}^{\dagger\dagger}$ &    0.27(5)             &    0.28(5)              &   0.26(3)                &   0.25(4)             &  0.2(1)      &       \\
{\it Gauss}$_4$ &E$_{l4}$, keV &     2.8(1)             &    2.80(5)              &   2.82(3)                &   2.8(1)             & 2.80(9)              & S~XVI $K_{\alpha}$ \\
             &  $N_{l4}^{\dagger\dagger}$ &    0.19(7)             &    0.20(7)              &   0.18(2)                &   0.17(9)             &  0.16(5)  &           \\
{\it Gauss}$_5$ &E$_{l5}$, keV &     1.82(9)             &    1.8(1)             &   1.83(4)                &   1.8(1)             & 1.83(8)              & Si~XIII\\
& &               &                 &                   &                &               & 
1$s^2~^1S$--1$s~2p^1P$\\
             &  $N_{l5}^{\dagger\dagger}$ &    0.21(5)             &    0.22(9)              &   0.20(5)                &   0.2(1)             &  0.21(9)  &           \\
{\it Gauss}$_6$ &E$_{l6}$, keV &     1.07(8)             &    1.05(6)              &   1.03(9)                &   1.0(1)             & 1.06(7)            & Ne~X 1$s$--2$p$/Fe~XVII \\
                           &                 &        &                                       &                                        &                                         &                                       
& 2$p^6~^1S$--2$p^5$4$d^1P$  \\
             &  $N_{l6}^{\dagger\dagger}$ &    0.14(4)             &    0.15(4)              &   0.13(2)                &   0.12(4)             &  0.11(3) &            \\
\hline
             &  $\chi^2_{red}$ (dof) &     0.86(464)           &    0.98(1271)              &  0.96(981)                 &    0.89(244)            &    0.96(189)            &\\
\hline
      \end{tabular}
    \label{tab:fit_table_suzaku+asca_ss_cyg}
\\$^\dagger$ The spectral model is  {\it tbabs*(CompTB + Gaussian$_1$ + Gaussian$_2$ + Gaussian$_3$ + Gaussian$_4$ + Gaussian$_5$ + Gaussian$_6$)}, 
normalization parameter of  {\it CompTB} component is in units of $L_{35}^{\rm soft}/d^2_{10}$, 
where $L_{35}^{\rm soft}$ is 
the soft photon  luminosity in units of 10$^{35}$; 
$d_{10}$ is the distance to the source in units of 10 kpc 
and {\it Gaussian} components are in units of $10^{-2}\times$ total photons cm$^{-2}$s$^{-1}$ in line. 
\end{table*}

\begin{figure}
\centering
\includegraphics[scale=1.50,angle=0]{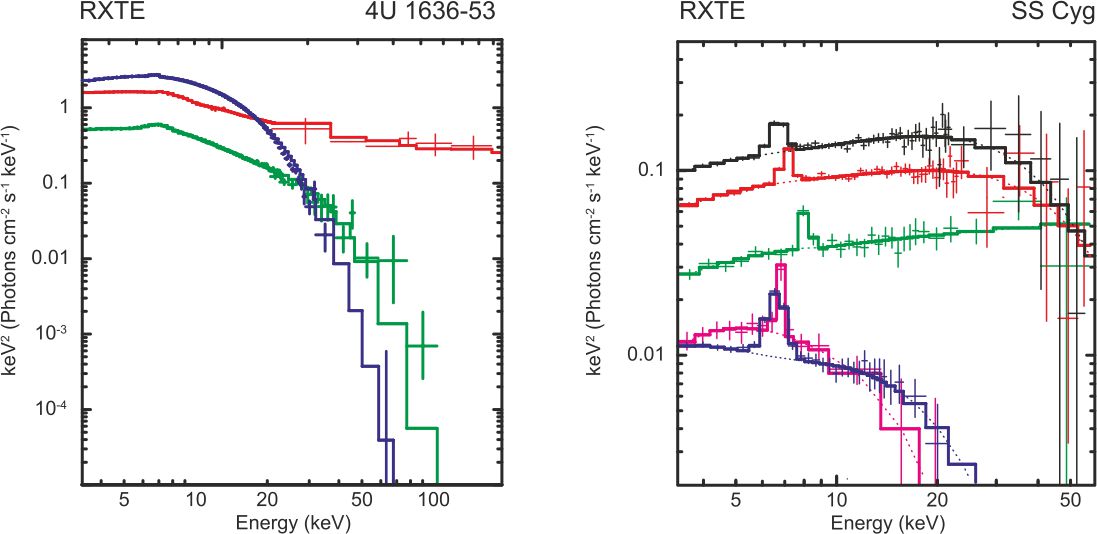}
\caption{Left: Four representative $E*F(E)$ spectral diagram of 4U~1636--53 during the LHS, IS and HSS events. Data are taken from RXTE observations 
10088-01-08-040 (red, LHS), 10088-01-09-02 (blue, IS),  and 94437-01-01-010 (green, HSS). 
Right: Five representative $E*F(E)$ spectral diagram of SS~Cygni during the LSS, LHS, IS and HSS events. Data are taken from RXTE observations 95421-01-01-05 (cyan, LSS), 4012-01-05-00 (blue, LSS),  95421-01-03-05 (green, LHS), 50011-01-69-00 (red, IS),  and 40012-01-15-00 (black, HSS). 
}
\label{rxte_sp_ev_ss_cyg}
\end{figure}

Five representative $E*F(E)$ spectral diagram of SS~Cygni during the LSS, LHS, IS and HSS events are presented in Fig.~\ref{rxte_sp_ev_ss_cyg}. Data are taken from {\it  RXTE} observations 95421-01-01-05 (cyan, LSS), 4012-01-05-00 (blue, LSS),  95421-01-03-05 (green, LHS), 50011-01-69-00 (red, IS),  and 40012-01-15-00 (black, HSS). 

Although {\it RXTE} data do not have this capability in terms of line resolution  for 4U1636--35 and SS~Cygni, they cover a larger time interval and allow us to study the long-term evolution of source properties and  make the complete picture of which is shown in the next section.

\subsection{Overall pattern of X-ray properties versus transient behavior 
\label{evolution_11}}

\subsubsection{Evolution of X-ray  spectral properties during transitions in 4U~1636--53 \label{evolution_NS}}


As was mentioned above, at time scales longer thah one minute, 4U~1636--53 exhibits two kinds of variability,  slow and mild. The former one  (slow) has a time scale of order of a few years. This slow variability is seen in the faint  
and  bright  phases 
which are related to low (200 -- 2010) and high (1996 -- 2000) luminosities, respectively. On the other hand the mild variability has a time scale of order of days and modulation depth is typically 20\% in the 3 -- 10 keV flux. The ASM (2 -- 12 keV) mean flux correlates with the {\it CompTB} normalization ($N_{com}$) and tends to anti-correlate with the electron plasma temperature of the Compton cloud, 
$T_e$ (Fig.~\ref{lc_1998}). Such mild variability is detected for both, the faint and bright phases  
for 4U~1636--53. It should be noted that the {\it CompTB} normalization, $N_{com}$ changes   larger in the   bright  state than that  during the  faint state.

One can relate slow and mild variabilities of 4U~1636--53 in order to demonstrate   slow and mild changes of the mass accretion rate, respectively. The slow variability by factor five has been seen in the 1996 -- 2010 observations by ASM/{\it RXTE}.  
We found that  the X-ray spectra of  4U~1636--53 over the bright and faint phases  are quite stable. 
The {faint/bright phase} transitions (see Fig.~\ref{lc_1998}) are characterized  by the  spectra with almost constant photon index $\Gamma$ about 2. We have also  established  common characteristics of the  rise-decay spectral transition  of 4U~1636--53 based on their  spectral parameter evolution of X-ray emission  in  the energy range from 3 to 50 keV using {\it RXTE}/PCA data.  
It is important to emphasize  that  normalization of the thermal  component is a factor of  2 higher in the bright phase than that in the faint phase, although  the photon indices $\Gamma$ for each of these spectra only vary in the range  from 1.9 to 2.1 and  are  concentrated  around $\Gamma$=2 (see  Fig. \ref{gam_Te}).

A number of X-ray flaring episodes of 4U~1636--53 has been  detected with {\it RXTE} during  1996 -- 2010 ($R1 - R5$ sets) with a good rise-decay coverage.  We have searched for  common spectral and timing features which can be revealed during these spectral transition episodes.  We present the combined results of  the spectral analysis of these observations  using  our  spectral model,  {\it tbabs*(Blackbody +  CompTB + Gaussian)}.  in   
{\it RXTE}/ASM  count rate is shown on the top panel of  Figure  \ref{lc_1998}. Further, from the  top to the bottom,  we show a change of the  transition layer  electron temperature $kT_e$. One can clearly see the {low amplitude} spectral transition { on time scales of $\sim$ 1 -- 2 days} from the bright phase 
to the faint phase 
at  MJD 51570/51552 while the electron temperature, $kT_e$ only varies from 2.3/2.5 keV to 20/40 keV during this transition,  respectively.  

Normalization of  the {\it CompTB} ({green} points) 
is shown in the next panel  of Fig.~\ref{lc_1998}. In particular, one can see from this figure  how the {\it CompTB} normalization $N_{com}$ correlates with  
the variations of  ASM count rate and  the model flux in 3--10 keV energy band. 
The Comptonization  fraction, $f$  varies from 0.1 to 0.9 as seen next panel (blue points) 
while the index $\Gamma$  slightly varies with time around 2 (see Fig. \ref{lc_1998}). 
 Note that 
in most  cases the soft disk  radiation  of 4U~1636--53  is subjected to reprocessing in a Compton cloud  and only some fraction of 
disk emission component ($1-f$) is directly seen by the Earth observer. 
The energy spectrum of 4U~1636--53 during almost all states is dominated by a Comptonized component 
while  the direct disk emission is 
always weak and detectable in the flaring episodes  only  (see Fig.~\ref{lc_1998})

It is worth noting here that in a BH  the  spectral transition  related to a change of  the photon index $\Gamma$ (see e.g. ST09).  However, there is no a  one-to-one correspondence between $\Gamma$  and cutoff  (or efold) energy $E_{fold}$. Titarchuk \& Shaposhnikov (2010) 
 demonstrate using {\it RXTE} data for BH binary  XTE~J1550--564 that   $E_{fold}$ decreases when $\Gamma$ increases from 1.4 to $2.1-2.2$ until    $\Gamma$ reaches 2.2   and then   $E_{fold}$ increases \cite{ts10}.  
Thus {for a BH the main parameter used for the spectral transition definition    is a { variable} photon index, $\Gamma$ which monotonically increases when the source goes into the bright phase}.
In contrast, for NS and WD binaries, this approach does not work and another parameter is needed to track source outbursts, for example, normalization $N_{com}$ or electron temperature $T_e$ of the Compton cloud (TL) during the transition from a weak phase to a bright one.

%
%

We define  the 
state transition in a NS source  
in terms of the {\it CompTB} normalization. 
 In the NS case  we relate 
the  faint phase to
a low  normalization value  while 
the bright phase 
relate to a high  normalization value. 
 The {\it CompTB} normalization and electron temperature $T_e$ are a directly measurable quantity (Fig.~\ref{lc_1998}). In Figure~\ref{gam_Te} (left panel) we demonstrate dependence of the photon index $\Gamma$ on the electron temperature $kT_e$ using these best-fit parameters for 4U~1636--53 obtained with analysis of {\it Beppo}SAX, ASCA and {\it RXTE} data.  Furthermore, from  Figure~\ref{lc_1998} 
one can clear see  a monotonic behavior  of  $N_{com}$ vs $kT_e$ when the electron temperature, $kT_e$ decreases and  the soft flux  increases. 

%
%
\begin{figure}
\centering
\includegraphics[scale=0.6,angle=0]{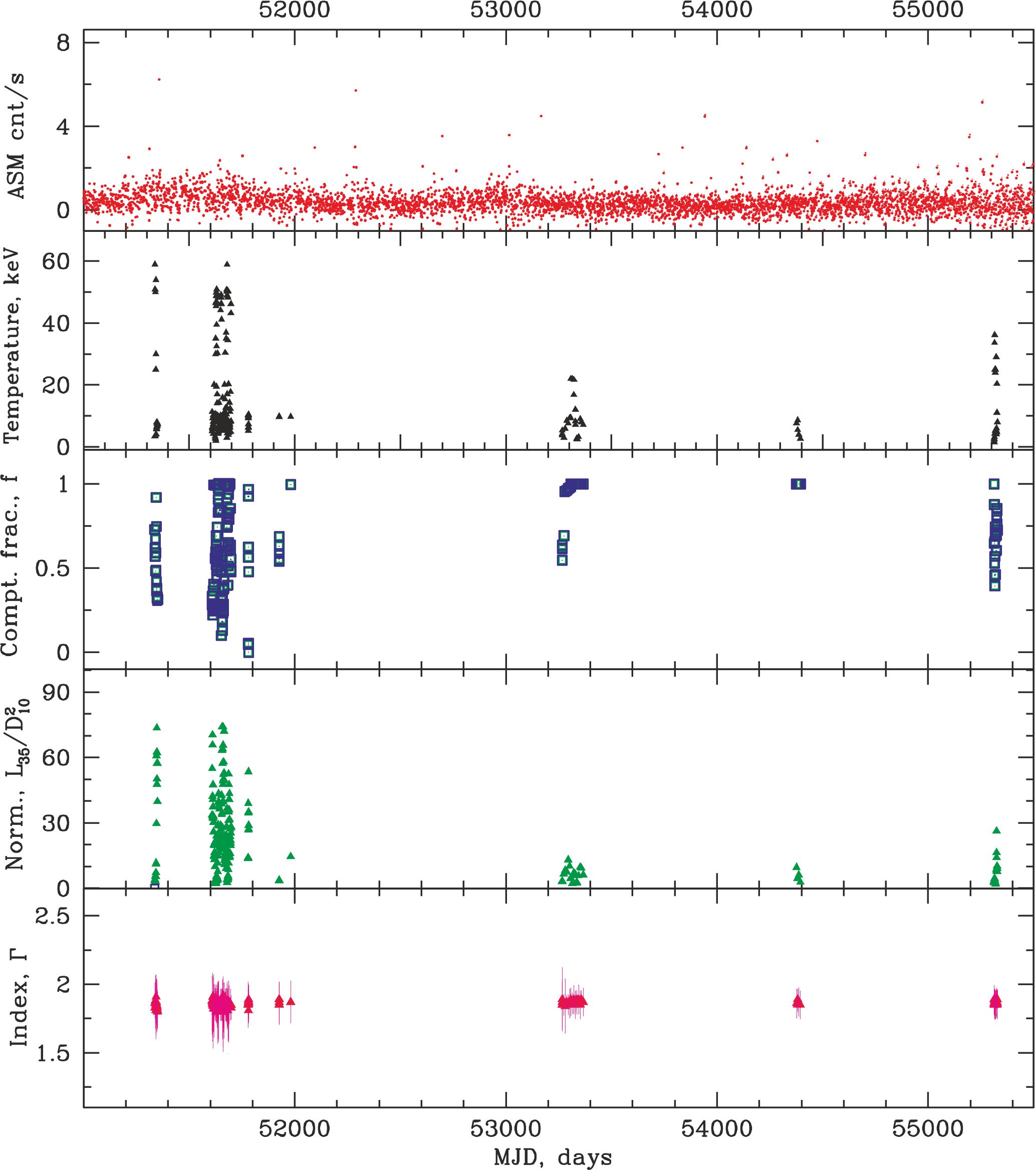}
\caption{{\it From Top to Bottom:}
Evolutions of the {\it RXTE}/ASM count rate (red), the electron temperature $kT_e$ (black),  {\it CompTB} normalization  ({\it blue}) the photon index $\Gamma$ ({\it crimson}) during 1996 -- 2010 outburst transitions set ({\it Rs1 -- Rs5}) for WD binary SS~Cyg. 
}
\label{frg_ss_cyg}
\end{figure}

\subsubsection{Evolution of X-ray  spectral properties during transitions in SS~Cygni \label{evolution_WD}}

In Figure~\ref{frg_ss_cyg}, we present an evolution of the X-ray parameters of SS~Cygni from 1996 to 2010.  The top panel shows the behavior of the ASM/{\it RXTE} count rate (red points), accompanied by frequent weak X-ray bursts [2--12 keV] at the level of 0--2 cnt/s and rare powerful bursts up to 6 cnt/s.  The electron temperature $T_e$ (the panel next to the top, black points) varies from 3 to 60 keV. At these moments, the Comptonization fraction $f$ (middle panel, blue points) ranges from 0.1 to 1, but tends to 1 more often, especially at the burst maximum. This  indicates a high degree of illumination of the transition layer. The normalization $N_{com}$ (penultimate panel below, green points) increases during outbursts from 0.1 to 75 $L_{35}/d^2_{10}$. However, the photon index $\Gamma$ (bottom panel, crimson points) remains nearly constant around $\Gamma=$ 1.85 for all states during the outburst.

\section{DISCUSSION \label{discussion}} 

\subsection{Comparison of timing properties between NS 4U~1636--53 and WD SS~Cygni 
sources
\label{timing analysis}}

We extracted light curves of 1 s resolution in the 2--4.5 keV energy range from all available proportional counter units (PCUs). The RXTE light curves were analyzed using the {\it powspec} task. The data analysis of the PDSs was performed using a simplified version of the diffusion model \cite{tsa07,ts08}
in which the PDS continuum shape at frequencies below the driving frequency can be approximated by an empirical KING model $P_X\sim (1.0 + (x/x_{\dagger})^2)$. 
Following \cite{tsa07}, the break frequency found in the PDS is related to a diffusion time of perturbation propagation while the QPO low frequency is an eigenfrequency of the volume (magnetoacoustic) oscillation of the medium (in our case it is a CC). We revealed that in 4U~1636--53 the mHz QPOs are consistent with variation of the plasma temperature of CC (or corona) around a NS. Namely, the anticorrelation of mHz-QPO frequency with the electron temperature indicates on the corona size is compacted when the $T_e$ is decreased. This allow us to associate mHz-QPOs origine with the corona dynamics during outburst cycle (Fig.~\ref{fig:mHz_QPO_T_e}, left).


It is now well known that dwarf novae exhibit different types of short-term variations~\cite{HackLaDawes93}:
QPOs (0.005 $<\nu_{QPO}<$ 0.05 Hz), coherent oscillations ($\nu\ge$ 0.1 Hz) and flickering ($\nu\simeq$ 0.01 Hz)~\cite{balman12,Maiolino20}.  
%
In the {\it right} panel of Fig.~\ref{fig:mHz_QPO_T_e}, we present the dependence of the index $\Gamma$  on the frequency of QPOs associated with the change of spectral states between LHS and HSS during outburst for NS (4U~1636--53, 4U~1728--34, 4U~1608--52, Aql~X--1, 4U~1705--44, GX~17+2 and Cyg~X--2) and WD (SS~Cygni). One can immediately see the difference between the NS and BC localization regions in this diagram: the NS are focused in the horizontal rectangular region for $1.9<\Gamma<2.1$ and $0.5 Hz<\nu_{QPO}<90 Hz$ (marked by a rectangle with a blue fill), although WDs are concentrated in a horizontal rectangular regions for $1.8<\Gamma<1.9$ and $0.005 Hz<\nu_{QPO}<0.05 Hz$. To complete the picture, we have added a similar dependence for BH (MAXI~J1348--630, XTE~J1550--564, GX~339--4, GRO~J1655-40, 4U~1543--47 and H~1743--322). It is clear that BHs are very different in terms of this diagram, showing oblique tracks over a wide range of QPO frequencies.

%
%
\begin{figure}
\centering
\includegraphics[scale=.9,angle=0]{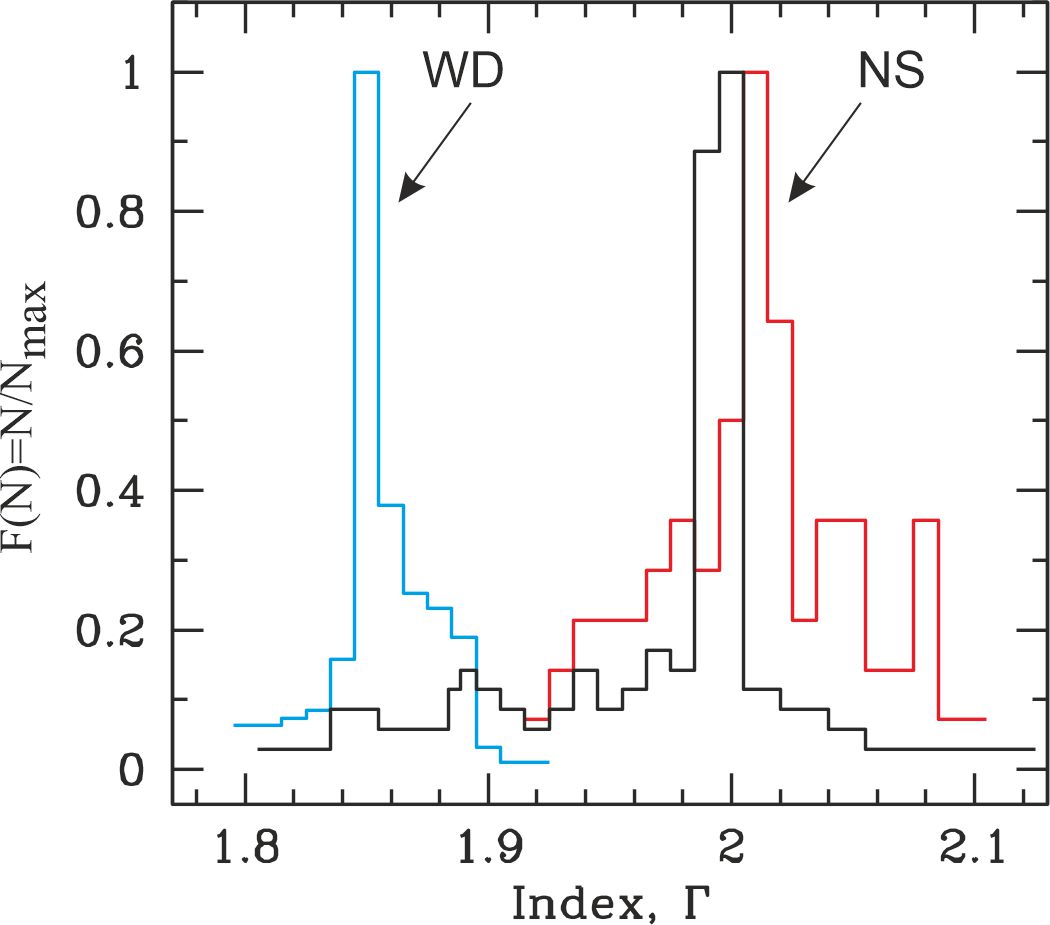}
\caption{Hystogram $N_{norm}=F(\Gamma)$ as a function  
of the best-fit photon power-law index $\Gamma$ obtained using a model {\it tbabs * (Blackbody +  CompTB + Gaussian)} for RXTE data of  WD SS~Cyg ({\it blue}) 
as well as NSs 4U~1636--53 ({\it red}) and 4U~1728--34 ({\it black}).  
Date for 4U~1728--34 was taken from \cite{st11}. The difference between WD and NS in terms of $N_{norm}=F(\Gamma)$  diagram peak, reflecting often occurence of $\Gamma$, are well seen: $\Gamma_{WD}$ focuses on 1.85, while $\Gamma_{NS}$ concentrates to 2.
}
\label{Hyst_NS_WD_index}
\end{figure}


\begin{figure}
\centering
\includegraphics[width=0.95\textwidth]{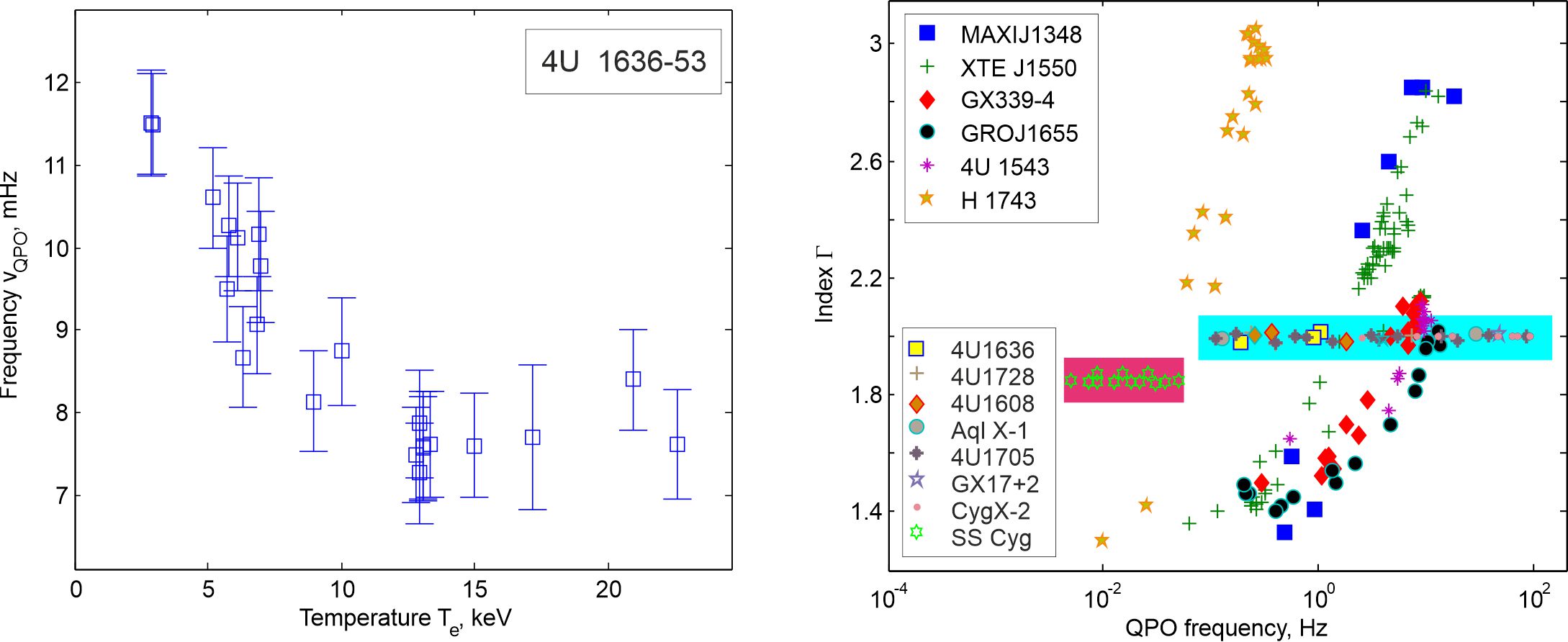}
\caption{\label{fig:mHz_QPO_T_e}{\it Left}: QPO frequency $\nu_{QPO}$ versus plasma temperature $T_e$ during outburst in 4U~1636--53 ($ R3$). {\it Right}: Photon index plotted versus QPO frequency for NS (4U~1636--53, 4U~1728--34, 4U~1608--52, Aql~X--1, 4U~1705--44, GX~17+2 and Cyg~X--2) and WD (SS~Cygni) sources and BHs (MAXI~J1348--630, XTE~J1550--564, GX~339--4, GRO~J1655-40, 4U~1543--47 and H~1743--322) during outburst transitions. NS are focused in a horizontal rectangular area for 1.9$<\Gamma<$2.1 and 0.5Hz~$<\nu_{QPO}<$~90Hz (marked with a blue filled rectangle), while WDs are concentrated in a horizontal rectangular area for 1.8$<\Gamma<$1.9 and 0.005Hz~$<\nu_{QPO}<$~0.05Hz. BHs show oblique tracks over a wide range of QPO frequencies. Data was taken from \cite{Sonbas22,DiSalvo01} (for NSs), \cite{balman12} (for WD) and \cite{st09,ST23} (for BHs).
}
\end{figure}

\subsection{Comparison of spectral properties between  {\it  NS} 4U~1636--53  and  {\it WD} SS~Cygni  sources \label{comparison_NS_WD}} 

In this section we 
compare 4U 1636--53 
with other NS binaries  and to identify differences and similarities  of these sources. 
We  make  a comparison 
with four NS 
sources 
(GX~340+0  \cite{STF13},  
4U~1728--34 \cite{st11}, 
GX~3+1 \cite{st12} 
 4U~1820--30  \cite{TSF13})  and three WD sources (U Gem, VW Hyi and SS Aur) using  the same spectral model,  which consists of the {\it Comptonized} continuum and the {\it Gaussian} line components. 
All parameters are summarized in Table~\ref{tab:fit_table_comb}, from which it can be seen that  all sources, regardless of distance, show the following behavior pattern:\\
\indent {$\bullet$}   the photon index (3$^{rd}$ column) is quasi-constant versus the electron temperature (4$^{th}$ column) for both source classes, but with fair different $\Gamma$-levels, namely with $\Gamma_{NS}\sim 2$ (for NSs) and $\Gamma_{WD}\sim 1.85$ (for WDs);\\
\indent {$\bullet$}  the range of the Comptonization fraction parameter  $f$ (7$^{th}$ column) is quite wide for both NS and WD sources (${\bar f} \sim 0.2-0.9$);\\
\indent {$\bullet$}  the plasma temperature $kT_e$ varies approximately in the same intervals $3<{\bar T_e}<30$ keV; and \\
\indent {$\bullet$}  the seed photon temperature $kT_s$ of NSs turns out to be higher (>1 keV) than for WDs (<0.2 keV).\\

Apparently, this fact indicates the difference in the surface temperatures of the central object ($T_s^{NS}\simeq 1.5$ keV and $T_s^{WD}\simeq 0.2$ keV). In this context, the high temperature of the NS provides a high reflectivity of the NS surface, while the low temperature of the WD ensures the absorption of radiation, which also affects the value of the photon index.

\subsection{Comparison with other results  in the literature}

Changes in the properties of the 4U~1636--53 continuum spectrum as the source moves through the CCD are generally consistent with those for other LMXB NSs. For example, \cite{ft11} measured the spectral index and electron temperature $kT_e$ of the Comptonizing component in LMXB NSs Sco~X--1, GX~349+2, X~1658--298, 1E~1724--3045, GX~17+2, Cyg~X--2, GX~340+0, GX~3+1 and GS~1826--238, and found that the spectral index $\alpha$ ($\Gamma-1$ in the {\it NthComp} component) remains more or less constant at $1\pm 0.2$. Using the {\it RXTE} and BeppoSAX observations, \cite{st11, st12} showed that in GX~3+1 and 4U~1728--34 the photon index $\Gamma$ also remains more or less constant during abrupt changes in the coronal temperature $T_e$. Titarchuk et al. (2013) found that in another NS binary, 4U~1820--30, the photon index remains almost constant for different source states \cite{TSF13}. Lyu et al. (2014) showed that the power index in 4U~1636--53 changes from $\sim$1.7 in the hard state to $\sim$2.8 \cite{Lyu14} in the soft state (depending on the model used to fit the iron line), which is somewhat larger than variations in sources studied by \cite{ft11}. Titarchuk et al. (2013) concluded that the spectral index quasi-stability is an intrinsic property of binary neutron stars, which is fundamentally different from the property of black hole binaries \cite{TSF13}. Seifina \& Titarchuk (2011) suggested that this index stability occurs when the energy released in the corona itself is much higher than the energy flux  from the disk \cite{st11}.
The continuum changes are also broadly consistent with the LMXB truncated disk scenario \cite{done07}. The temperature of the disk increases and the inner radius of the disk decreases as the estimated mass accretion rate increases. The electron temperature of the {\it NthComp} component and the surface temperature of the NS in \cite{Lyu14} are also consistent with this scenario.

Previous studies has shown that the parameters of the spectral components differ in different spectral states. The temperature of thermal components is usually in the range of 0.5--2.0 keV in the soft state \cite{DiSalvo00,Ooster01} and below 1 keV in the hard state \cite{Church01,Barret03}. The coronal temperature changes in the opposite way, from 2--3 keV in the soft state to several tens of keV in the hard state \cite{Gierlinski+Done03}.

Lyu et al. (2014) found that the temperature at the inner radius of the disk  in  4U 1636--53 increases significantly from $\sim$0.2 keV \cite{Lyu14} in the hard state to about 0.8 keV in the soft state (see their Fig. 4, top panel), while the {\it Blackbody} temperature increases slightly from $\sim$1.5 to $\sim$2 keV with increasing X-ray luminosity of the source (Fig. 4, second panel from the top). This is in full agreement with our results, taking into account the small differences between the models and the different meanings of model component bindings. So in our case, $kT_{BB}$ varies from 0.2 keV (in the hard state) to 0.7 keV (in the soft state), and we associate this temperature with the disk and its internal parts. Further, the temperature of the seed photons of the disk penetrating into the transition layer and participating in the Comptonization processes, $kT_s$, increases from 1.2 keV to 1.7 keV with the development of the outburst. This Lyu's {\it Blackbody} temperature, in fact, was described using a separate {\it blackbdody} component. We used the {CompTB} model, which allows us to determine this temperature of the seed photons $kT_s$ and immediately takes into account that these photons will participate in the processes of energy exchange with the hot electrons of the transition layer. For this reason, we do not use a separate {\it BBody} component, but the meaning of $kT_s$ is the same as the $kT_{BB}$ temperature used in \cite{Lyu14}. However, that their model  to fit the Comptonization component is different from the one we used here

Our spectral analysis results for SS~Cygni from the ASCA data are somewhat different from the results obtained in \cite{Done_Osborne97}. They described the observed spectrum of the X-ray continuum with a phenomenological model, a power law with a photon index of $\Gamma=$ 2, while fixing $\Gamma$ at this value. But we did not fix it and described the observed spectrum using a physically justified first principles model taking into account the interaction of matter and radiation in the X-ray range, the Comptonization model {\it CompTB} and obtained $\Gamma\sim$ 1.8. With taking into account these differences in the models, as well as observational errors, our and Done \& Osbotn's results  agree with each other.

%
%
\begin{table*}
  \centering 
 \caption{Comparisons of the best-fit parameters  of NS binaries 
(4U~1636--53, GX~340+0$^1$, 
 GX~3+1$^2$, 4U~1728--34$^3$, 4U~1820--30$^4$] 
with WD binaries (SS~Cyg, VW~Hyi, SS~Aur, U~Gem \cite{Maiolino20}).} 
 \begin{tabular}{llccccccc}
\hline\hline
Source &  Distance, & Index, &  $kT_e$,  & $ N_{\rm com}$, &  $kT_{s}$,  & $f$ \\
  name & kpc         &  $\Gamma$  &      keV         &  $L_{39}^{\rm soft}/{D^2_{10}}$        & keV  &  \\
\hline
GX 340+0     & 10.5$^{5}$  &  2.13$\pm$0.08    & 3--21     &  0.08--0.2  & 1.1--1.5 & 0.01--0.5   \\
GX 3+1       & 4.5$^{6}$  &  2.02$\pm$0.05       & 2.3--4.5 &  0.04--0.15 & 1.16--1.7 & 0.2--0.9   \\
4U~1728--34    & 4.2-6.4$^{7}$ &  1.99$\pm$0.02 & 2.5--15  & 0.02--0.09  &  1.3 & 0.5-1\\     
4U~1820--30  &    5.8-8$^{8}$  &  2.04$\pm$0.07  & 2.9--21  & 0.02--0.14  &  1.1--1.7 & 0.2--1\\     
4U~1636--53   & 7.4$^{9}$  &  2.00$\pm$0.02     & 2.7--30  &  0.01--0.08  & 1.1--1.8 & 0.1--1   \\
\hline
SS~Cyg  &  0.17$^{10}$  & 1.85$\pm$0.03 &    2.7--60  &  0.001--0.007     & 0.02--0.2$^{10}$   & 0.1--1   \\
VW~Hyi   & 0.05$^{10}$  & 1.75$\pm$0.09 &    5--10     &  0.0001--0.0006 & 0.1--0.15$^{11}$   & 0.1--1   \\
SS~Aur   &  0.17$^{10}$  & 1.85$\pm$0.04 &    6.1--37  &  0.0001--0.0005  & 0.05--0.17$^{10}$ & 0.1--1  \\
U~Gem    & 0.1$^{10}$   & 1.80$\pm$0.09 &    5.0--43  &  0.0001--0.0008  & 0.08--0.15$^{10}$ & 0.1--1  \\
\hline
      \end{tabular}
    \label{tab:fit_table_comb}
\\ 
$^{1}$ \cite{STF13}; 
$^{2}$ \cite{st12}; 
$^{3}$ \cite{st11};   
$^{4}$ \cite{TSF13}; 
$^{5}$ 
   \cite{Fender+Henry00,FvK98,chsw97}; 
$^{6}$ \cite{kk00,Ford00}; 
$^{7}$ \cite{par78};  
$^{8}$ \cite{ST04};  
$^{9}$ \cite{Ritter_Kolb03}; 
$^{10}$  \cite{Maiolino20}; 
$^{11}$ \cite{Nakaniwa19}. 
\end{table*}

\section{CONCLUSIONS}

We presented the results of spectral and timing analysis of 4U~1636--53 and SS~Cygni during transitions between the faint and bright phases. 
We analyzed all transient episodes for these sources and established clear observational 
differences between NS and WD during outbursts in these binaries using results from the {\it Beppo}SAX, Suzaku, ASCA  and {\it RXTE} missions. 
We show that the X-ray broad-band energy spectra for both sources during all spectral states can be well fitted by a   sum of  the  Comptonization and   {\it Gaussian} components. For the NS 4U~1636--35, we added {\it Blackbody } component to better fit the source spectrum. A wide variation of  parameter  $f=0.1-0.9$, obtained in the framework of our spectral model,   points out a significant  variation of the illumination of  the Comptonization region (transition layer)  by  X-ray soft photons  in  4U~1636--53 and SS~Cygni.    
 We  also show  that the spectral index $\alpha$ of the best-fit Comptonized component  for 4U~1636--53 is almost constant, about 1 and consequently almost  independent of   {\it CompTB} normalization $N_{com}$ 
(which is proportional to the disk mass accretion rate $\dot m$) and the electron temperature $T_e$. 
Using the plasma temperature $T_e$ as an outburst tracer, we confirmed that the X-ray spectral index of SS~Cygni is also constant throughout outbursts, but it tends to different value, $\alpha\to 0.85$. This is consistent with what was previously found from less extensive data from {\it RXTE}, Chandra and XMM-Newton for a number of other  low-mass X-ray NSs and WDs. We  should remind a reader  that this index stability ($\alpha_{NS}\to 1$)  has been recently suggested for a wide variety of other NS sources, for example,  for  {\it atoll} sources: GX~3+1 \cite{st12}, X~1658--298, GS~1826--238, 1E~1724--3045, 4U~1728--34 \cite{ft11,st11}  and also for  Z-sources: Cyg~X--2, Sco~X--1, GX~17+2, GX~340+0, GX 349+2 , which were  observed by {\it Beppo}SAX and {\it RXTE} \cite{ft11,st11}. In turn, recall also that such index stability ($\alpha_{WD}\to 0.85$) has recently been suggested for a number of other WD sources, for example, for U~Gem, VW~Hyi and SS~Aur, which were observed by XMM-{\it Newton} and {\it Chandra}. We interpret this difference in terms of the TL model, 
which is sensitive to different boundary conditions and the temperatures of NS and WD surfaces. In fact, the higher temperature obtained for the NS surface provides a higher reflectivity than for the WD surface. As a consequence, the WD surface absorbs X-rays produced in the TL better than the NS surface, which leads to different values of the spectral index. 

In addition,  we revealed that in 4U~1636--53 the mHz-
QPOs are consistent with variation of the plasma temperature of Compton cloud (or corona) around a NS.  Namely, the anticorrelation of mHz-QPO frequency with the electron temperature indicates on the corona size is compacted when the $T_e$ is decreased. This indicates on the corona size is compacted when the $T_e$ is decreased and allow us to associate mHz-QPOs origine with the corona dynamics during outburst cycle. We also stated different $\Gamma-\nu_{QPO}$ patterns for NS and WD in terms of the highest QPO frequency  component of the signal above the Poisson noise in the power spectra. NS X-ray spectra most often show quasi-stable behavior at the $\Gamma\to 2$ level and are characterized by LF-QPO in a wide frequency range (0.5--90 Hz), although WD X-rays stabilize at the $\Gamma\to 1.85$ level and  accompanied with QPOs in a narrower frequency range (0.005--0.05 Hz). This index effect, now well established for 4U~1636--53 and WD SS~Cygni in extensive observations, was previously found in other low-mass X-ray NSs and WDs. The index constancy for WD and its lower value in comparison with NS in combination with a higher QPO frequency indicates a fundamental difference between NS and WD sources and is a new set of features that is easily detected in observations.

\section*{ACKNOWLEDGMENTS}
The Authors are very  happy to get a careful reading  and editing our  manuscript by Chris Shrader.  






\bibliographystyle{Frontiers-Vancouver} 

\begin{thebibliography}{100}
\bibitem[Sonbas et al. (2022)]{Sonbas22} Sonbas E, Mohamed K, Dhuga KS, et al. A Temporal Scale to Track the Spectral Transitions in Low-Mass X-ray Binaries. MNRAS. 2022;511:2535-2543.
\bibitem[Mohamed et al. (2021)]{Mohamed21} 
Mohamed K, Sonbas E, Dhuga KS, et al. A minimal timescale for the continuum in 4U~1608-52 and Aql X-1. MNRAS. 2021;502:L72-L78.
\bibitem[Hasinger \& van der Klis (1989)]{hasinger89}
Hasinger G, van der Klis M. Two patterns of correlated X-ray timing and spectral behaviour in low-mass X-ray binaries. A\&A. 1989;225:79-96.
\bibitem[Liu et al. (2007)]{liu07}
Liu QZ, van Paradijs J, van den Heuvel EPJ. A catalogue of low-mass X-ray binaries in the Galaxy, LMC, and SMC (Fourth edition). AAP. 2007;469:807-810.
\bibitem[Fiocchi et al. (2006)]{fiocchi06}
Fiocchi M, Bazzano A, Ubertini P, et al. Disk-Jet Coupling in the Low-Mass X-Ray Binary 4U 1636-53 from INTEGRAL Observations. ApJ. 2006;651:416-420.
\bibitem[Corbel \& Kaaret (2000)]{CK00}
Corbel S., Kaaret P, Simultaneous ASCA and RXTE observations of the Atoll X-Ray Binary 4U1636-536. Rossi2000: Astrophysics with the Rossi X-ray Timing Explorer. March 22--24, 2000 at NASA's Goddard Space Flight Center, Greenbelt, MD USA. 2000;6.
\bibitem[Smale \& Mukai (1988)]{smale_mukai88}
Smale AP, Mukai K. The orbital variability of XB 1636-536. MNRAS. 1988;231:663-671.
 \bibitem[van Paradijs et al. (1990)]{van_Paradijs90}
van Paradijs J, van der Klis M, van Amerongen, 
et al. The orbital period of 4U/MXB 1636-53 (V801 Arae). A\&A. 1990;234:181.
\bibitem[Casares et al., (2006)]{Casares06}
Casares J, Cornelisse R, Steeghs D, et al. 
Detection of the irradiated donor in the LMXBs 4U 1636-536 (=V801 Ara) and 4U 1735-444 (=V926 Sco). MNRAS. 2006;373:1235-1244.
\bibitem[Strohmayer (1998)]{stroh98}
 Strohmayer T. RXTE observations of GX 3+1. AIP Conf. Proc., 1997 October Astrophysics Conference in Maryland: "Accretion Processes in Astrophysics, Some Like it Hot".
1998;431;397-400.
\bibitem[Giles et al. (2002)]{Giles02} 
Giles AB, Hill KM, Strohmayer TE, et al. 
Burst Oscillation Periods from 4U 1636-53: A Constraint on the Binary Doppler Modulation. ApJ. 2002;568:279-288.
\bibitem[Bitner et al. (2007)]{Bitner07} 
Bitner MA, Robinson EL, Behr BB. The Masses and Evolutionary State of the Stars in the Dwarf Nova SS Cygni.  ApJ. 2007;662:564-573.

\bibitem[Galloway et al. (2006)]{Galloway06} 
Galloway DK, Psaltis D, Muno MP, et al. 
Eddington-limited X-Ray Bursts as Distance Indicators. II. Possible Compositional Effects in Bursts from 4U 1636-536. ApJ. 2006;639:1033-1038.
\bibitem[Galloway et al. (2018)]{Galloway18}
Galloway DK, in't Zand JJM, Chenevez J, et al. The Influence of Stellar Spin on Ignition of Thermonuclear Runaways. ApJ. 2018;857:L24. 
\bibitem[Galloway et al. (2020)]{Galloway20}
Galloway DK, in't Zand J, Chenevez J, et al. The Multi-INstrument Burst ARchive (MINBAR). ApJS. 2020;249:32.
\bibitem[Sanna et al. (2013)]{Sanna13} 
Sanna A, Hiemstra B, M$\bar e$ndez, M, et al. 
Broad iron line in the fast spinning neutron-star system 4U 1636-53. MNRAS. 2013;432:1144-1161.
\bibitem[Frank et al. (1987)]{Frank87} 
Frank J, King AR, Lasota J-P. The light curves of low-mass X-ray binaries. A\&A. 1987;178:137-142.
\bibitem[Lyu et al. (2014)]{Lyu14}
Lyu M, M$\bar e$ndez M, Sanna A, et al. 
Iron-line and continuum variations in the XMM-Newton and Suzaku spectra of the neutron-star low-mass X-ray binary 4U 1636-53. MNRAS. 2014;440:1165-1178.
\bibitem[Shih et al. (2005)]{Shih05}
Shih IC, Bird AJ, Charles PA, et al. 
Periodic variability during the X-ray decline of 4U 1636-53. MNRAS. 2005;361:602-606.
\bibitem[Belloni et al. (2007)]{belloni07} 
Belloni T, Homan J, Motta S, et al. 
Rossi XTE monitoring of 4U1636-53 -- I. Long-term evolution and kHz quasi-periodic oscillations. MNRAS. 2007;379:247-252.
\bibitem[Strohmayer et al. (1996)]{Strohmayer96}
Strohmayer TE, Zhang W, Swank JH, et al. 
Millisecond X-Ray Variability from an Accreting Neutron Star System. ApJ. 1996;469:L9.
\bibitem[van der Klis et al. (1996)]{van_der_Klis96}
van der Klis M, Swank JH, Zhang W, et al. 
 Discovery of Submillisecond Quasi-periodic Oscillations in the X-Ray Flux of Scorpius X-1. ApJ. 1996;469:L1.
\bibitem[Mendez et al. (2006)]{Mendez06}
M$\grave e$ndez M. On the maximum amplitude and coherence of the kilohertz quasi-periodic oscillations in low-mass X-ray binaries. MNRAS. 2006;371:1925-1938.

\bibitem{Revnivtsev01}
Revnivtsev M, Churazov E, Gilfanov M, Sunyaev R. New class of low frequency QPOs: Signature of nuclear burning or accretion disk instabilities? A\&A. 2001;372:138-144.
\bibitem{Fei21}
Fei Z, Lyu M, Méndez M, et al 
The Harmonic Component of the Millihertz Quasi-periodic Oscillations in 4U 1636-53.  ApJ. 2021;922:119. 
\bibitem{Heger07}
Heger A, Cumming A, Woosley SE. Millihertz Quasi-periodic Oscillations from Marginally Stable Nuclear Burning on an Accreting Neutron Star. ApJ. 2007;665:1311-1320.
\bibitem{Keek14}
Keek L, Cyburt RH, Heger A. Reaction Rate and Composition Dependence of the Stability of Thermonuclear Burning on Accreting Neutron Stars. ApJ.  2014;787:101. 
\bibitem{Yu+vanderKlis02}
Yu W, van der Klis M. Kilohertz Quasi-periodic Oscillation Frequency Anticorrelated with Millihertz Quasi-periodic Oscillation Flux in 4U 1608-52. ApJL. 2002;567:L67-L70.
\bibitem{Altamirano08}
Altamirano D, van der Klis M, Wijnands R, et al. 
Millihertz Oscillation Frequency Drift Predicts the Occurrence of Type I X-Ray Bursts. ApJL. 2008;673:L35.
\bibitem{Mancuso19}
Mancuso GC, Altamirano D, García F, et al. Discovery of millihertz quasi-periodic oscillations in the X-ray binary EXO 0748-676. MNRAS. 2019;486:L74-L79.
\bibitem{Linares12}
Linares M, Altamirano D, Chakrabarty D, et al. 
Millihertz Quasi-periodic Oscillations and Thermonuclear Bursts from Terzan 5: A Showcase of Burning Regimes. ApJ. 2012;748:82. 
\bibitem{Lyu15}
Lyu M, Méndez M, Zhang G, et al. 
Spectral and timing analysis of the mHz QPOs in the neutron-star low-mass X-ray binary 4U 1636-53. MNRAS. 2015;454:541-549.
\bibitem{Yu+vanderKlis02}
Yu W, van der Klis M. Kilohertz Quasi-periodic Oscillation Frequency Anticorrelated with Millihertz Quasi-periodic Oscillation Flux in 4U 1608-52. ApJL. 2002;567:L67-L70.
\bibitem{Stiele16}
Stiele H, Yu W, Kong AKH. Millihertz Quasi-periodic Oscillations in 4U 1636-536: Putting Possible Constraints on the Neutron Star Size. ApJ. 2016;831:34. 
\bibitem{Strohmayer18}
Strohmayer TE, Gendreau KC, Altamirano D, et al. NICER Discovers mHz Oscillations in the “Clocked” Burster GS 1826-238. ApJ. 2018;865:63. 
\bibitem[Barret  (2001)]{Barret01}Barret D. The broad band x-ray/hard x-ray spectra of accreting neutron stars. Adv. Space Res.  2001;28:307-321.
\bibitem[Lin et al. (2007)]{Lin07}
Lin D, Remillard RA, Homan J. Evaluating Spectral Models and the X-Ray States of Neutron Star X-Ray Transients. ApJ. 2007;667:1073-1086.
\bibitem[Lewin \& van der Klis (2006)]{Lewin+vanderKlis06}
Lewin WHG, van der Klis M. Compact Stellar X-Ray Sources. Cambridge, UK:: Cambridge Univ. Press). 2006;39.

\bibitem[Titarchuk et al. (2014)]{TSS14}  
Titarchuk L, Seifina E, Shrader C. X-Ray Spectral and Timing Behavior of Scorpius X-1. Spectral Hardening during the Flaring Branch. ApJ. 2014;789:98. 
\bibitem[Friend et al. (1990)]{Friend90}
Friend MT, Martin JS, Smith RC, Jones DHP. The 8190-A sodium doublet in cataclysmic variables III. Too cool for credibility. MNRAS. 1990;246:654-667.
\bibitem[Ritter \& Kolb (2003)]{Ritter_Kolb03}
Ritter H, Kolb U. Catalogue of cataclysmic binaries, low-mass X-ray binaries and related objects (Seventh edition). A\&A. 2003;404:301-303.
\bibitem[Harrison et al. (2004)]{Harrison04} 
Harrison TE, Johnson JJ, McArthur BE, et al. An Astrometric Calibration of the MV-Porb Relationship for Cataclysmic Variables based on Hubble Space Telescope Fine Guidance Sensor Parallaxes.  AJ. 2004;127:460-468.
\bibitem[Mauche \& Robinson (2001)]{Mauche+Robinson01} 
Mauche CW, Robinson EL. First Simultaneous Optical and Extreme-Ultraviolet Observations of the Quasi-coherent Oscillations of SS Cygni. ApJ. 2001;562:508-514.
\bibitem[Ishida et al. (2009)]{Ishida09} 
Ishida M, Okada S,  Hayashi T, et al. 
Suzaku Observations of SS Cygni in Quiescence and Outburst. PASJ. 2009;61:77-91.
\bibitem{balman12} 
Balman S, Revnivtsev M. X-ray variations in the innner accretionflow of dwarf novae.' A\&A. 2012;546:A112. 
\bibitem{F08}
Farinelli R, Titarchuk L, Paizis A, Frontera F. A New Comptonization Model for Weakly Magnetized, Accreting Neutron Stars in Low-Mass X-Ray Binaries. ApJ. 2008;680:602-614. 
\bibitem[Mitsuda et al. (1984)]{Mitsuda84}
Mitsuda K, Inoue H, Koyama K, et al. Energy spectra of low-mass binary X-ray sources observed from Tenma. PASJ. 1984;36:741-759.
 \bibitem[Shaposhnikov \& Titarchuk (2009)]{st09}
Shaposhnikov N, Titarchuk L. Determination of black hole masses in galactic black hole binaries using scaling of spectral and variability characteristics, ApJ, 2009;699:453-468. 
\bibitem[Titarchuk \& Seifina (2023)]{ST23}
Titarchuk L, Seifina E. MAXI~J1348--630: Estimating the black hole mass and binary inclination using a scaling technique". A\&A. 2023;669:A57. 
\bibitem[Makishima et al. (1986)]{Makishima86}
Makishima K, Maejima Y, Mitsuda K, et al. 
Simultaneous X-Ray and Optical Observations of GX 339-4 in an X-Ray High State. ApJ. 1986;308:635-643.
\bibitem[Zdziarski et al. (1996)]{Zdziarski96}
Zdziarski AA, Johnson WN, Magdziarz P. Broad-band $\gamma$-ray and X-ray spectra of NGC 4151 and their implications for physical processes and geometry. MNRAS. 1996;283:193-206.
\bibitem[Zycki et al. (1999)]{Zycki99}
Zycki PT, Done C, Smith DA. The 1989 May outburst of the soft X-ray transient GS 2023+338 (V404 Cyg). MNRAS. 1999;309:561-575.
\bibitem[Madejski et al. (1999)]{Madejski99}
Madejski GM, Sikora M, Jaffe T, et al. X-Ray Observations of BL Lacertae during the 1997 Outburst and Association with Quasar-like Characteristics. ApJ. 1999;521:145-154.
\bibitem[Shakura \& Sunyaev  (1973)]{ss73} 
Shakura NI, Sunyaev RA. Black holes in binary systems. Observational appearance. A\&A. 1973;24:337-355.
\bibitem[Balman et al. (2001)]{balman01}
Balman $\bar S$, Godon P, Sion EM, et al. XMM-Newton Observations of the Dwarf Nova RU Peg in Quiescence: Probe of the Boundary Layer.
ApJ. 2011;741:84. 
\bibitem[Chenevez  et al. (2006)]{ch06}
Chenevez J, Falanga M, Brandt S, et al. Two-phase X-ray burst from GX 3+1 observed by INTEGRAL. A\&A. 2006;449:L5-L8.
\bibitem{TS10}
Titarchuk L, Shaposhnikov N. Implication of the Observed Spectral Cutoff Energy Evolution in XTE J1550-564. ApJ. 2010;724:1147-1152.
\bibitem{st11}
Seifina E, Titarchuk L. On the Constancy of the Photon Index of X-Ray Spectra of 4U 1728-34 through All Spectral States. ApJ. 2011;738:128-148
\bibitem{tsa07}
Titarchuk L, Shaposhnikov N, Arefiev V. Power Spectra of Black Holes and Neutron Stars as a Probe of Hydrodynamic Structure of the Source: Diffusion Theory and Its Application to Cygnus X-1 and Cygnus X-2 X-Ray Observations. ApJ. 2007;660:556-579.
\bibitem{ts08}
Titarchuk L, Shaposhnikov N. On the Nature of the Variability Power Decay toward Soft Spectral States in X-Ray Binaries: Case Study in Cygnus X-1. ApJ. 2008;678:1230-1236.
\bibitem[Titarchuk \& Shaposhnikov  (2010)]{ts10} 
Titarchuk L,  Shaposhnikov N. Implication of the Observed Spectral Cutoff Energy Evolution in XTE J1550-564. ApJ. 2010;724:1147-1152.
\bibitem{HackLaDawes93}
Hack M, Ladous C, Jordan SD, et al. Cataclysmic variables and related objects. NASSP. 1993;507.
\bibitem[Done et al. (2007)]{done07}
Done C, Gierl$\bar i$nski M, Kubota A. Modelling the behaviour of accretion flows in X-ray binaries. Everything you always wanted to know about accretion but were afraid to ask. 
  A\&AR. 2007;15:1-66.
\bibitem[Maiolino  et al. (2020)]{Maiolino20} 
Maiolino T, Titarchuk L, D'Amico F. et al. Testing Comptonization as the Origin of the Continuum in Nonmagnetic Cataclysmic Variables: The Photon Index of X-Ray Emission. ApJ. 2020;900:153. 
\bibitem[Seifina et al.  (2013)]{STF13}
Seifina E, Titarchuk L, Frontera F. Stability of the Photon Indices in Z-source GX 340+0 for Spectral States. ApJ. 2013;766:63. 
\bibitem[Seifina \& Titarchuk  (2012)]{st12}
Seifina E, Titarchuk L. GX 3+1: The Stability of Spectral Index as a Function of Mass Accretion Rate. ApJ. 2012;747:99. 
\bibitem[Titarchuk et al. (2013)]{TSF13}
  Titarchuk L, Seifina E, Frontera F. Spectral State Evolution of 4U 1820-30: The Stability of the Spectral Index of the Comptonization Tail. ApJ. 2013;767:160. 
\bibitem[Fender \& Hendry (2000)]{Fender+Henry00}
Fender RP,  Hendry MA. The radio luminosity of persistent X-ray binaries.  MNRAS. 2000;317:1-8.
\bibitem[Ford \& van der Klis (1998)]{FvK98}
Ford E,  van der Klis M. Strong Correlation between Noise Features at Low Frequency and the KilohertzQuasi-Periodic Oscillations in the X-Ray Binary 4U 1728-34. ApJ. 1998;506:L39-L42.
\bibitem[Christian  \& Swank, 2006]{chsw97}
Christian DJ, Swank JH. The Survey of Low-Mass X-Ray Binaries with the Einstein Observatory Solid-State Spectrometer and Monitor Proportional Counter. ApJS. 1997;109:177-224.
\bibitem[Kuulkers \& van der Klis (2000)]{kk00}
Kuulkers E, van der Klis M. The first radius-expansion X-ray burst from GX 3+1.  A\&A. 2000;356:L45-L48.
\bibitem{Ford00}
Ford EC, van der Klis M. Mendez, M., et al. Simultaneous Measurements of X-Ray Luminosity and Kilohertz Quasi-Periodic Oscillations in Low-Mass X-Ray Binaries. ApJ. 2000;537:368-373.
\bibitem[van Paradijs  (1978)]{par78} 
 van Paradijs J. Average properties of X-ray burst sources. Nature. 1978;274:650-653.
 \bibitem[Shaposhnikov \& Titarchuk (2004)]{ST04}
Shaposhniko N, Titarchuk L. On the Nature of the Flux Variability during an Expansion Stage of a Type I X-Ray Burst: Constraints on Neutron Star Parameters for 4U 1820-30. ApJL. 2004;606:L57-L60.
\bibitem[Nakaniwa et al. (2019)]{Nakaniwa19}
Nakaniwa N, Hayashi T, Takeo M, et al. 
Variation of mass accretion rate on to the white dwarf in the dwarf nova VW Hyi in quiescence. MNRAS. 2019;488:5104-5113.
\bibitem[Farinelli \& Titarchuk (2011)]{ft11}
Farinelli R, Titarchuk L. On the stability of the thermal Comptonization index in neutron star low-mass X-ray binaries in their different spectral states.  A\&A. 2011;525:102. 
\bibitem[Done \& Osborne (1997)]{Done_Osborne97}
Done C, Osborne JP. The X-ray spectrum of the dwarf nova SS CYG in quiescence and outburst. MNRAS. 1997;288:649-664.
\bibitem{DiSalvo01}
Di Salvo T,Mendez M,van der Klis M, et al. 
Study of the Temporal Behavior of 4U 1728–34 as a Function of Its Position in the Color-Color Diagram. ApJ. 2001;546:1107-1120. 
\bibitem[Oosterbroek et al. (2001)]{Ooster01}
Oosterbroek T, Barret D, Guainazzi M, et al. 
. Simultaneous BeppoSAX and RXTE observations of the X-ray burst sources GX 3+1 and Ser X-1. A\&A. 2001;366:138-145.
\bibitem[Church et al. (2001)]{Church01}
Church MJ, Baluci$\bar n$ska-Church M. Results of a LMXB survey: Variation in the height of the neutron star blackbody emission region. A\&A. 2001;369:915-924.
\bibitem[Barret et al. (2003)]{Barret03}
Barret D, Olive JF, Oosterbroek T. Simultaneous BeppoSAX and Rossi X-ray timing explorer observations of 4U 1812-12. A\&A. 2003;400:643-647.
\bibitem[Gierlinski \& Done (2003)]{Gierlinski+Done03}
Gierli$\bar n$ski M, \& Done C. The X-ray/$\gamma$-ray spectrum of XTE J1550-564 in the very high state. MNRAS. 2003;342:1083-1092.
\bibitem[DiSalvo et al. (2000)]{DiSalvo00}
Di Salvo T, Iaria R, Burderi L, et al. 
The Broadband Spectrum of MXB 1728-34 Observed by BeppoSAX.  ApJ. 2000;542:1034-1040.
\bibitem{Balman20} Balman S. Accretion Flows in Nonmagnetic White Dwarf Binaries as Observed in X-rays. Advances in Space Research. 2020;66:1097-1122.
\bibitem[Okada  et al. (2008)]{Okada08} 
\bibitem[Mukai et al. (2003)]{Mukai03}
Mukai K, Kinkhabwala A, Peterson JR, et al. 
Two Types of X-Ray Spectra in Cataclysmic Variables.  ApJL. 2003;586:L77-L80.
Okada S, Nakamura R, Ishida M. Chandra HETG Line Spectroscopy of the Nonmagnetic Cataclysmic Variable SS Cygni. ApJ. 2008;680:695-704.

\bibitem[Byckling  et al. (2010)]{Byckling10}
Byckling K, Mukai K, Thorstensen JR, et al. 
Deriving an X-ray luminosity function of dwarf novae based on parallax measurements. MNRAS. 2010;408:2298-2311.
\bibitem[Xu et al. (2016)]{Xu16} 
Xu X-j, Wang QD, Li X-D. Fe Line Diagnostics of Cataclysmic Variables and Galactic Ridge X-Ray Emission. ApJ. 2016;818:136. 
\bibitem[Kimura et al. (2021)]{Kimura21}      
Kimura M, Yamada S, Nakaniwa N, et al. On the nature of the anomalous event in 2021 in the dwarf nova SS Cygni and its multi-wavelength transition. PASJ. 2021;73:1262-1279.
\bibitem[Parmar et al. (1997)]{parmar97} 
Parmar AN,  Martin DDE, Bavdaz M, et al. The low-energy concentrator spectrometer on-board the BeppoSAX X-ray astronomy satellite. A\&AS. 1997;122:309-326.
\bibitem[Boella et al. (1997)]{boel97} 
Boella  G, Chiappetti L, Conti G, et al. The medium-energy concentrator spectrometer on board the BeppoSAX X-ray astronomy satellite.  A\&AS. 1997;122:327-340.
\bibitem[Frontera et al. (1997)]{fron97}
Frontera F, Costa E, dal Fiume D, et al. PDS experiment on board the BeppoSAX satellite: design and in-flight performance results. SPIE.  1997;3114:206-215.
\bibitem[Tanaka et al.  (1994)]{Tanaka94} 
Tanaka Y, Inoue H, Holt SS. The Fluorescence-Dominated X-Ray Spectrum of the Spiral Galaxy NGC 6552. PASJ. 1994;46:L141-L146.
\bibitem[Mitsuda et al. (2007)]{Mitsuda07}
Mitsuda K, Bautz M, Inoue H. et al. The X-Ray Observatory Suzaku. PASJ. 2007;59:S1-S7.
\bibitem{Koyama07}
Koyama K, Tsunemi H, Dotani T, et al. X-Ray Imaging Spectrometer (XIS) on Board Suzaku. PASJ. 2007;59:23-33.
\bibitem[Gaia Collaboration, Brown et al. (2018)]{Gaia18} 
Gaia Collaboration, Brown AGA, Vallenari A, Prusti T, et al. Gaia Data Release 2. Summary of the contents and survey properties. A\&A.  2018;616:A1. 

\end{thebibliography}

\end{document}